\documentclass[twocolumn,prb,showpacs,preprintnumbers,amsmath,amssymb]{revtex4}
\usepackage{color}
\usepackage[usenames]{xcolor}
\usepackage[normalem]{ulem}
\newcommand{\bea}{\begin{eqnarray} }
\newcommand{\eea}{\end{eqnarray}}
\newcommand{\bean}{\begin{eqnarray*}}
\newcommand{\eean}{\end{eqnarray*}}
\newcommand{\nn}{\nonumber \\}
\def\bfg#1{{\mbox{\boldmath $#1$}}}

\def\B{{\bf B}}

\def\J{{\bf J}}
\def\jac{{\cal J}}
\def\j{{\bf j}}
\def\k{{\bf k}}

\def\r{{\bf r}}

\def\Kb {{\bfg \kappa}}
\def\zxi {{\bfg \xi}}

\def\btimes{~{\bf \times}~}
\def\bnabla{{\bf \nabla}}
\def\bcdot{~{\bf \cdot}~}
\newcommand{\lbs}{\left (}
\newcommand{\rbs}{\right )}
\newcommand{\lbm}{\left\lbrack}
\newcommand{\rbm}{\right\rbrack}
\newcommand{\lbl}{\left \{ }
\newcommand{\rbl}{\right \} }

\def\av#1{\left\langle #1 \right\rangle}
\def\od#1,#2{\frac{d#1}{d#2}}
\def\odz#1,#2{\frac{d^2#1}{d{#2}^2}}
\def\pd#1,#2{\frac{\partial #1}{\partial #2}}
\def\pdz#1,#2{\frac{\partial^2 #1}{\partial {#2}^2}}

\def\pdd#1,#2{\frac{\partial^3 #1}{\partial {#2}^3}}
\def\pdv#1,#2{\frac{\partial^4 #1}{\partial {#2}^4}}
\def\pdzz#1,#2,#3{\frac{\partial^2 #1}{\partial {#2}\partial{#3}}}

\def\eq#1{Eq.~(\ref{#1})}
\def\eqn#1{(\ref{#1})}
\usepackage{graphicx}
\usepackage{dcolumn}
\usepackage{bm}

\begin{document}

\bibliographystyle{unsrt}
%
\title{X point effects on      the ideal MHD modes in tokamaks
\\in the description of dual-poloidal-region safety factor}
%
%
\author{Linjin  Zheng,\footnote{Corresponding author: Linjin Zheng, \\email: lzheng@austin.utexas.edu} M. T. Kotschenreuther, F. L. Waelbroeck, and M. E. Austin}
\affiliation{Institute for Fusion Studies,
University of Texas at Austin,
Austin, TX 78712}
\date{\today}

\begin{abstract}

  The flux coordinates with dual-region safety factor ($q$) 
  in the poloidal direction are developed in this work. 
The X-point effects on    the ideal MHD modes in tokamaks
 are then analyzed using
this coordinate system.   Since the X-point effects mainly affect the edge region, 
the modes localized at the tokamak edge are particularly examined.
Two types of modes are studied.
The first is related to the conventional peeling or peeling-ballooning modes.
The mode existence aligned with the local magnetic field 
in the poloidally core region as observed experimentally is confirmed. 
The X points are shown to contribute to a stabilizing effect   
for the conventionally treated modes with the surface-averaged $q$
and with the tokamak edge portion truncated. 
The other is the axisymmetric modes localized in the vicinity of X points,
which can affect the  cross-field-line transport near the X points. 
The existence of axisymmetric modes
points to the possibility of applying a toroidally axisymmetric resonant magnetic perturbation (RMP) in the X-point area for 
mitigating the edge localized modes, which  
 can be   an alternative to the current RMP design. 
   The dual $q$ description also has important implications for the existing non-axisymmetric 
RMP concept.
It helps to understand why the RMP suppression of edge localized modes is difficult to achieve
in the double-null tokamak configurations and points to the possibility  of further improving the current RMP concept by considering the alignment to the local $q$.


\end{abstract}


\maketitle

 \section{Introduction}

\label{sec.1}

Nowadays, the high mode (H mode) confinement has become the standard scheme for conventional
tokamaks with positive triangularity.\cite{hmo} However,  the H mode confinement is often tied with 
the so-called edge localized modes (ELMs), that can discharge the pedestal heat to divertors.\cite{hmo}  
Such a discharge may severely damage the divertors. 
Theoretically, the peeling-ballooning modes have been accepted as the interpretation of ELMs,\cite{sny} 
    Nevertheless, the separatrix effects (or X-point effects) on the plasma edge stability
remain an active research subject. 
An important progress about the separatrix effects on the ballooning
modes has been  made in Ref. \onlinecite{bis1}. It has been employed to explain the  physics
picture of  tokamak H-mode confinement.\cite{bis2}
There are also other investigations based on the model equilibria,
for example in Refs. \onlinecite{myr} and \onlinecite{saa}. 
  References \onlinecite{webprl} and \onlinecite{web} also reported the 
 stabilization of a toroidal plasma's separatrix in magnetohydrodynamic (MHD) description.
Numerically, the   GATO and KINX codes are also developed
to compute the X-point effects with the finite element method.\cite{gato,kinx}  
  Both GATO and KINX have been routinely used to study the MHD stability
of the equilibria with X points, for example in Refs. \onlinecite{alan} and \onlinecite{med}.
    Different from the Fourier-decomposition-based
codes, the finite-element-based codes provide the possibility to
 address the issue of  high local safety factor $q$ (i.e., small field line pitch)
in the vicinity of X points. In this work, we introduce an alternative approach based on the dual $q$ coordinates
to treat this problem.
  Besides, as shown in the recent paper Ref. \onlinecite{xeq}, the plasma boundary 
near the X point can only be of the hyperbola type or in the X shape with the plasma segment in a right angle
(i.e., $90$ degrees).   
This shows that the equilibria used by some early studies of X-point effects need to be modified.
As will be seen, our current  theory, however, works with the Solov\'ev equilibrium\cite{sol} with the tiny thin edge
 layer truncated.   As will be seen, the main difference from
 the previous works lies in that the dual $q$ coordinates are used in the current work. 
 
\begin{figure}[h]\vspace*{2mm}
\hspace*{-2mm}
\includegraphics[width=55mm,angle=-90]{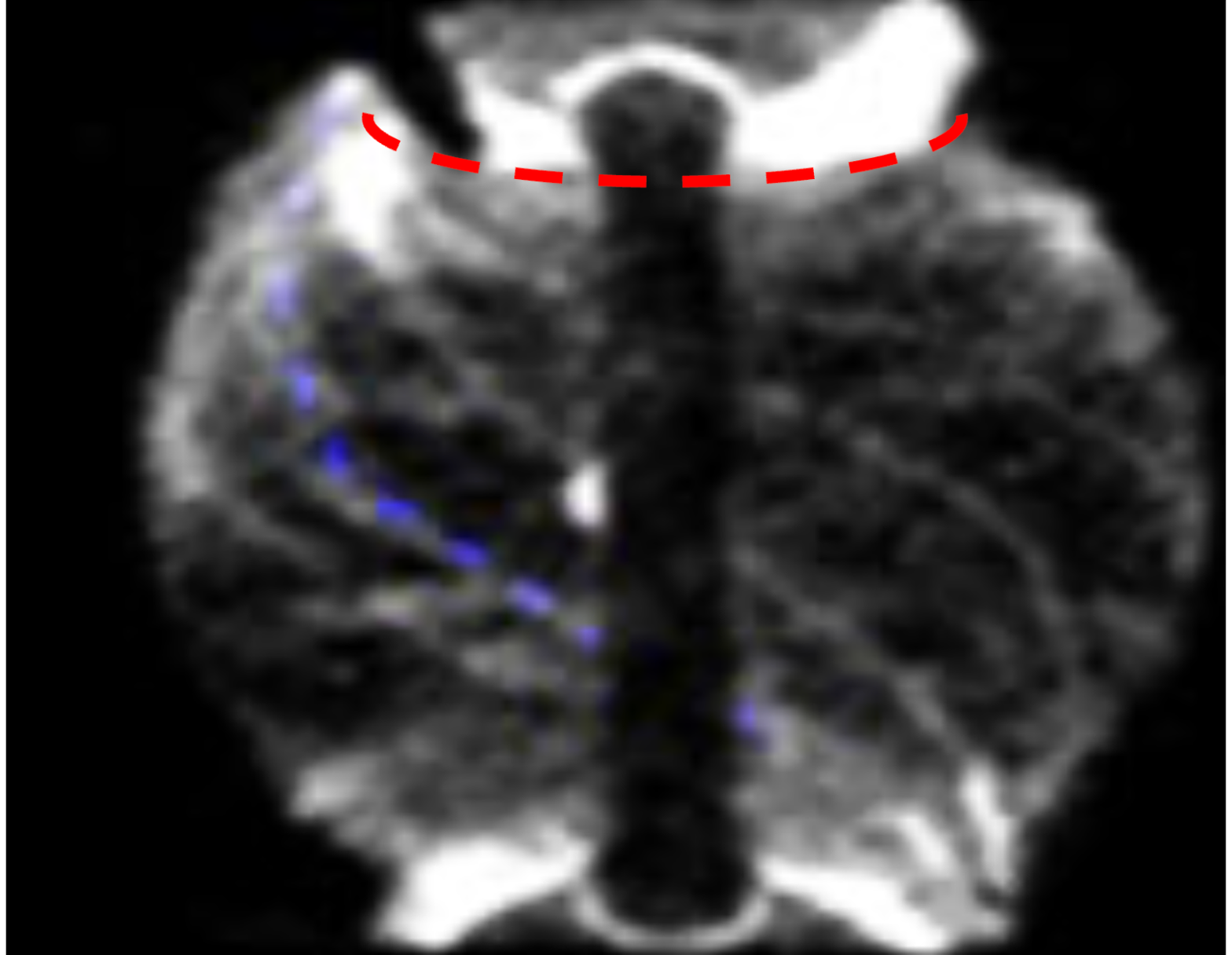}\vspace*{3mm}
\caption{The peeling ballooning filaments observed in the MAST experiment,   
The blue dashed and newly introduced  red dashed curves mark two typical field line 
pitches in  regions $\Theta_{core}$
and $\Theta_X$ respectively.  Reprinted with permission  from 
A. Kirk, et al.,
{\it Phys. Rev. Lett.} {\bf 96}, 185001 (2006),\cite{mast} Copyright (2006) by the
American Physical Society.}
\label{f1}
\end{figure}

In fact, in a tokamak, the poloidal magnetic field vanishes only at the X point. If one introduces the local 
safety factor, it tends to infinity only at the X points, while remaining finite elsewhere. It is  the surface average
  in the definition of $q$ in the conventional flux coordinates
that makes the safety factor tend to infinity everywhere on a surface  as approaching the plasma edge.
  This can be seen later on in the main text  in the flux coordinate representation of magnetic field in
\eq{bb} and the definition of $q$ in \eq{qq}.
Using the surface-averaged safety factor can be misleading in interpreting the relevant physics at 
the plasma edge. 
Experimentally, the MAST experimental observation as shown in Fig. \ref{f1} indicates that 
the perturbation filaments actually are aligned with the local magnetic field line, i.e., following the local $q$.\cite{mast,mastfig}  It, therefore, does not fit
the surface-averaged $q$ description.    In the surface-averaged $q$ description,
the poloidal and toroidal wave numbers, $k_\theta$ and $k_\psi$, for the modes aligned with
the local $q$ in the poloidally core region would become infinite
in the vicinity of X points.\cite{bis2,myr} This does not appear in Fig. \ref{f1}. 
Theoretically, in the surface-averaged  $q$ description, 
the Alfv\'en resonance condition 
 $m-nq=0$ requires that for a finite toroidal mode number $n$, the poloidal mode number
 $m$ has to be infinite due to $q\to\infty$. This implies that the perpendicular wavelength turns to vanish.
This certainly does not reflect experimental observations and is also  unacceptable physically,
 especially for MHD description. 
The perpendicular wavelength cannot be shorter than the Larmor radius   for the MHD formalism 
to be relevant.
Furthermore,    if the surface-averaged $q$ is used, the extremely large
magnetic shear appears everywhere on the last few closed flux surfaces, which again oversimplifies 
the equilibrium description. The magnetic shear only becomes extremely large in the vicinity of X point. 
There are consequences for this oversimplified description. For example,
 since the apparent mass effect is proportional to  the safety factor square, 
the surface-averaged $q$  leads the mass of parallel motion to become extremely large 
everywhere on the last few closed flux surfaces, and so are the finite Larmor radius effects.
All of these show that one should give up the single surface-averaged $q$ description
at the plasma edge. 

  Let us discuss further  the experimental observation.
 From physics consideration,
other than that the MHD perturbation follows the local field 
line as pointed out in Ref. \onlinecite{mast}, one would not expect otherwise unless the nonideal MHD
effects play a role. Noting that the actual $q$ is of dual regions in the poloidal direction,  one would not expect
the confinement of plasma to behave differently in the experiments. 
One may argue that the filament structure shown in Fig. \ref{f1} looks like the 
usual ballooning mode feature with  $k_\|\ll k_\bot$, where
$k_\|$ and  $k_\bot$ are respectively the parallel and perpendicular wave numbers. 
  Actually, it is not. 
The subtle physics picture of flux-tube-like  modes, like  the
 ballooning or interchange type of modes, needs to be considered. 
For the flux-tube-type modes, one needs to examine the 
 field line bundle inside the flux tube, instead of a single filed line pitch.
 In doing so, one see that the bundle shrinks sharply from the poloidally core region to the X point or expands dramatically from the X point to the poloidally core region. The flux tube modes tied to the field line bundle 
 with finite width in the perpendicular region in the poloidally core region (blue dashed curve in  Fig. \ref{f1})
 are fundamentally different from the flux tube modes tied to the field line bundle 
 with finite width in the perpendicular region in the vicinity of X points  (red dashed curve in Fig. \ref{f1}).
   This indicates there can be two primary modes: one aligns with the local field line pitch in the poloidally core region and the other follows the field lines in the vicinity of X point. 
  This picture goes beyond oversimplified considerations
 based on a single field line pitch without taking into account the field-line-bundle shrinkage or expansion.
This shows  that the     dual-poloidal-region $q$ description is more appropriate   in considering this
dual mode feature. 

\begin{figure}[h]\vspace*{-0mm}
\centering
\includegraphics[width=60mm,angle=0]{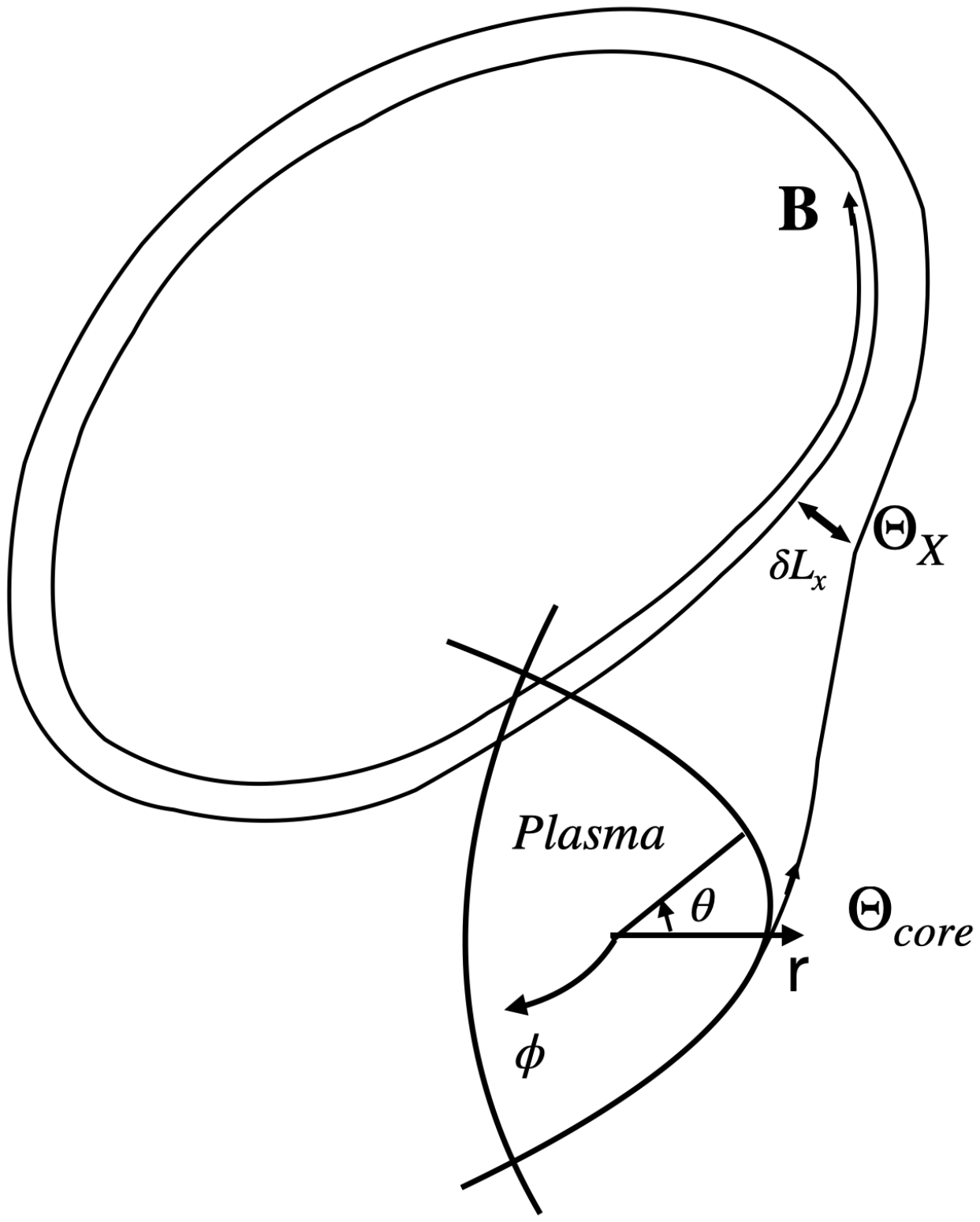}\vspace*{-3mm}
\caption{Schematic plot of the magnetic field line pattern in an equilibrium with X points.
The coordinate system ($r,\theta,\phi$) are shown, where $r$ is the minor radius
 from the magnetic axis, $\theta$ is the poloidal
angle, and $\phi$ denotes the axisymmetric toroidal angle.}
\label{f2}
\end{figure}

Let us also discuss further  the numerical treatment of X point effects. Leaving aside
the existence issue of the X point equilibrium as pointed out in Ref. \onlinecite{xeq},
the single surface-averaged $q$ description results in an infinite number of rational surfaces at plasm edge 
in the radial direction and the requirement of infinite poloidal Fourier harmonics 
(or highly dense poloidal grids)
to describe it. 
None of the existing MHD codes   based on the surface-averaged $q$ to
represent the magnetic field
can handle this situation. Even if it did, the distance between the rational surfaces or the perpendicular
wavelength of high $m$ modes would be well below the Larmor radius. 
This makes the MHD description inapplicable   in the vicinity of X point. 
To explain this, we introduce the schematic plot of magnetic field line pattern in Fig. \ref{f2}. 
In the inner and outer boards of plasma torus ($\Theta_{core}$) the field lines have a finite pitch. However,
in the vicinity of X point, the pitch tends to zero and the distance from the magnetic field line 
to itself after a circular turn starting from it, $\delta L_X$ in Fig. \ref{f2}, becomes very small as approaching
the X point. The MHD energy minimization can take the perpendicular wavelength to be less than 
$\delta L_X$. However, there is an applicability limitation of MHD description. When $\delta L_X$ is
less than the ion Larmor radius, one cannot use the MHD theory. Therefore, 
in the dual-poloidal-region $q$ description, the poloidally core region can be treated by the ideal MHD for
the peeling-ballooning type of modes   (or external kink modes), but the vicinity of X point $\Theta_X$  non-ideal MHD 
description is required. The requirement of nonideal MHD description is not just
for the dual-poloidal-region $q$ description. It is a general requirement for X-point physics.
  Note that the X-point singularity   effect on $q$ is spread over the magnetic surface 
 in the surface-averaged $q$ description and the ideal MHD 
 codes based on the surface-averaged $q$ requires to minimize the energy
in each interval between two neighboring rational surfaces.  This type of codes in principle
cannot even handle the poloidally core region because
the distance between the resonant surfaces is less
than the Larmor radius.  

To display the actual features of local $q$ in the poloidal direction,
 the Solov\'ev equilibrium is examined in this work.\cite{sol} 
 It indeed indicates that the local $q$ tends to infinity  only in the vicinity of X points.
 This leads us to develop the flux coordinates with dual-poloidal-region safety factor $q$ on a magnetic surface,
 which describes separately the local safety factor in the vicinity of X point ($\Theta_X$) and elsewhere in the
 poloidally core regions ($\Theta_{core}$).
 In this coordinate system, the infinite or large   local safety factor  appears only in the vicinity of
X points, while the   local safety factor remains finite elsewhere. 

With the dual-poloidal-region $q$ coordinates, the X-point effects on the   
MHD modes  at tokamak edge
are revisited in this work. 
  Since the X-point effects mainly affect the edge region, 
the modes localized at the tokamak edge are particularly examined. 
Two types of  magnetohydrodynamic  modes are studied in this description.
The first is related to the conventional peeling or peeling-ballooning modes   including
the external kink modes.
The presence of X points is   confirmed to contribute  a stabilizing effect
    for the conventionally treated modes  with the surface-averaged $q$
and with the tokamak edge portion truncated. 
The other is the n = 0 axisymmetric modes localized in the vicinity of X point,
which can affect the edge transport picture near the X points.
The former determines  the perturbation filaments as observed in MAST.\cite{mast} 
The latter can affect the cross field line transport in the divertor region.\cite{evan,rmp1,rmp}
It is pointed out that the existence of axisymmetric modes may be exploited to
mitigate the edge localized modes by applying simply axisymmetric resonance magnetic fields near 
the X points.

 The manuscript is organized as follows. In Sec. II, the safety factor features 
 in the Solov\'ev equilibria;  In Sec. III,
dual-poloidal-region safety factor coordinates are developed; In Sec. IV, the X-point effects on    the 
ideal MHD  modes
are investigated;   In Sec. V, 
the conclusions and discussion are presented.

 \section{Safety factor features in the Solov\'ev equilibria}

\label{sec.2}

To study the X-point effects on MHD modes, several equilibrium models have been used for example in 
Refs. \onlinecite{myr} and \onlinecite{saa}, as well
as the so-called $s-\alpha$ model.\cite{bis1}   In these models, the equilibria with X points
are constructed semi-analytically.
In this work, we use the Solov\'ev equilibria.\cite{sol}  
The  Solov\'ev solution can approximate the DIII-D-like cross-section. For our investigation,
the main concern is  the profile of local safety factor   (the local field line pitch) in the poloidal direction. 
This is mainly related to the Jacobian behavior near the X point. 
The Solov\'ev equilibrium solution is sufficient to explain the situation.

For axisymmetric tokamak configuration, the magnetic field can be expressed
as follows
\bea
\B&=& \bnabla\phi\btimes\bnabla\chi +f(\chi)\bnabla\phi
\nn 
&=& \bnabla\phi\btimes\bnabla\chi +q(\chi)\bnabla\chi\btimes\bnabla \theta_p.
\label{bb}
\eea
Here, $\chi$ is the poloidal magnetic flux, $\phi$ is the toroidal angle, 
 $f$ denotes the poloidal current density flux, $q(\chi)$ denotes the safety factor,
 and $\theta_p$ is the poloidal angle in the  so-called PEST coordinates,\cite{pest} which is 
 defined as follows
\bea
\theta_p&=& \frac f{q}\int_0^{\theta_{eq}} d\theta_{eq} \frac {{\cal J} }{X^2},
\nn
q&=& \frac f{2\pi}\oint d\theta_{eq} \frac {{\cal J} }{X^2}   \equiv  \frac f{2\pi}\oint d\theta_{eq} q_{local},
\label{qq}
\\
{\cal J} &=& \frac 1{ \bnabla\phi\btimes\bnabla\chi\bcdot \bnabla\theta_{eq}},
\label{jac}
\eea
Here, $\theta_{eq}$ is the  poloidal angle specified in the equilibrium code   and the local 
safety factor  $q_{local} = f {\cal J} /X^2$.

The Grad-Shafranov equation in cylindrical coordinates ($X,Z,\phi$)  is
\bean
X\pd, X\frac1X\pd \chi,X+\pdz \chi,Z& =& -\mu_0P_\chi' X^2- ff_\chi', 
\eean
where  $P$ is the pressure,
$X$ is the major radius, $Z$ is the height,   $\mu_0$ is the magnetic constant,
and the prime denotes the derivative with respect to $\chi$. In the Solov\'ev solution,
it is assumed that
\bean
 -\mu_0P_\chi' =a~~~ \hbox{and}~~~ -ff_\chi'=b  X_0^2,
 \eean
  where  $X_0$ is the major radius of magnetic axis.
The exact Solov\'ev equilibrium solution is then given as follows\cite{sol}
\bea
\chi &=&\lbm (b+c_0)X_0^2+c_0(X^2-X_0^2)\rbm \frac{Z^2}2 
\nn&&
+\frac18(a-c_0)(X^2-X_0^2)^2,
\label{chi0}
\eea
where    $a,b$, and $c_0$ are  constant parameters.

We first determine the separatrix of the solution in \eq{chi0}. This can be obtained by the
stationary points of $\chi$:
\bean
\pd \chi, Z&=&0 ~~\to~~~ X^2-X_0^2 = -\frac{b+c_0}{c_0}X_0^2,
\nn
\pd \chi, {(X^2-X_0^2)}&=&0 ~~\to~~~ Z^2=-\frac12\frac{a-c_0}{c_0} (X^2-X_0^2). 
\eean
Therefore, given the X-point coordinates $(X_s,Z_s)$, one can determine the parameters
\bean
\frac a{c_0}=1-2\frac{Z^2_s}{ X_s^2-X_0^2},
~~~
\frac b{c_0}&=& -1 - \frac{X_s^2-X_0^2}{X_0^2}.
\eean

\begin{figure}[h]\vspace*{-18mm}
\centering
\includegraphics[width=80mm]{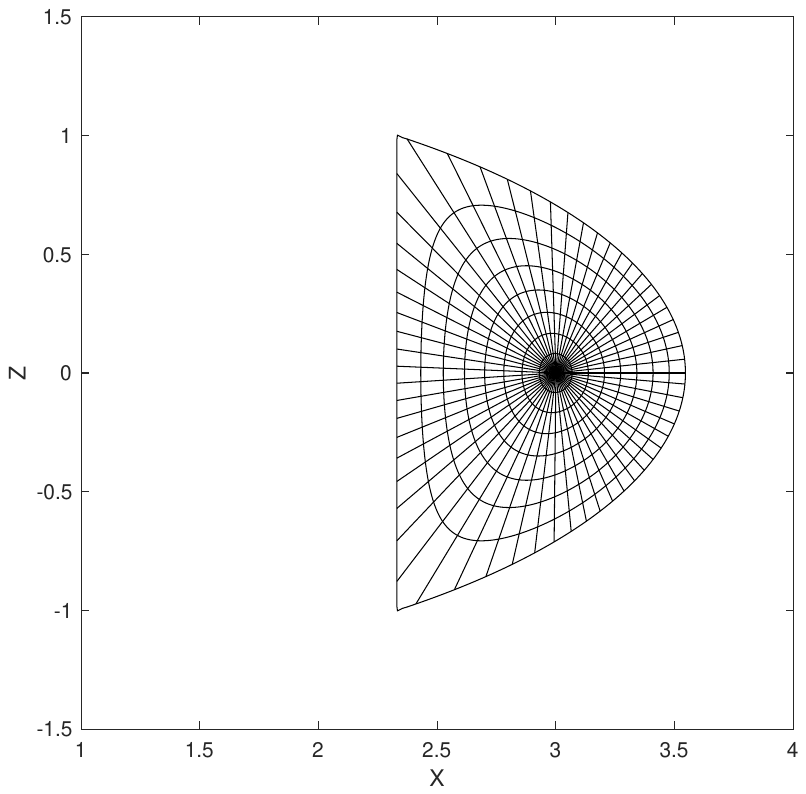}\vspace*{-19mm}
\caption{The   equilibrium cross section   of the Solov\'ev solution.}
\label{f3}
\end{figure}

\begin{figure}[h]\vspace*{-14mm}
\centering
\includegraphics[width=70mm,angle=-0]{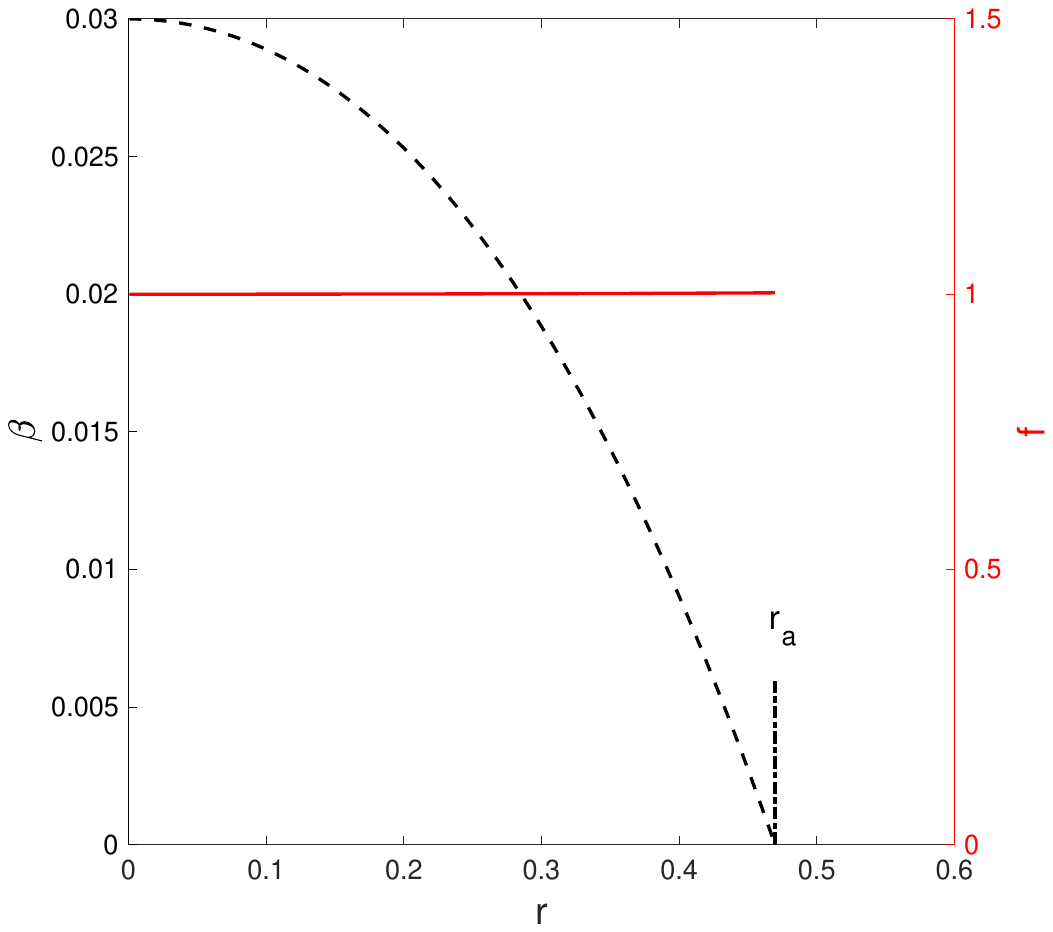}\vspace*{-13mm}
\caption{The    local $\beta$ (black, dashed) and $f$ (red, solid) profiles versus the minor radius ($r$) on the mid-plane
  on the out board side for the    configuration as described 
in Fig. \ref{f3}.   Here, $r_a$ indicates the plasma boundary.}
\label{f4}
\end{figure}

To be specific, we choose   $X_0=3$, $X_{s}=2.33$,  $Z_{s}=1$, and the beta at the magnetic axis  $\beta_0=0.03$. 
The  equilibrium cross sections are plotted  in Fig. \ref{f3}, with the $\beta$ and $f$ profiles
given in Fig. \ref{f4}.  From \eq{chi0} one can see that $c_0^2$ can be absorbed into
the definitions of $a$ and $b$. We therefore choose $c_0=1$ for simplicity. 
  In this case,    $a= 1.5601$ and $b=-0.6032$. 
Here, it should be pointed out that in determining $f$ from $b$ there is an integration constant,
which is actually related to the magnitude of toroidal field, The integration constant is therefore
used to scale the beta value at the magnetic axis   as in Ref. \onlinecite{bate}.

\begin{figure}[htp]\vspace*{-18mm}
\centering
\includegraphics[width=80mm]{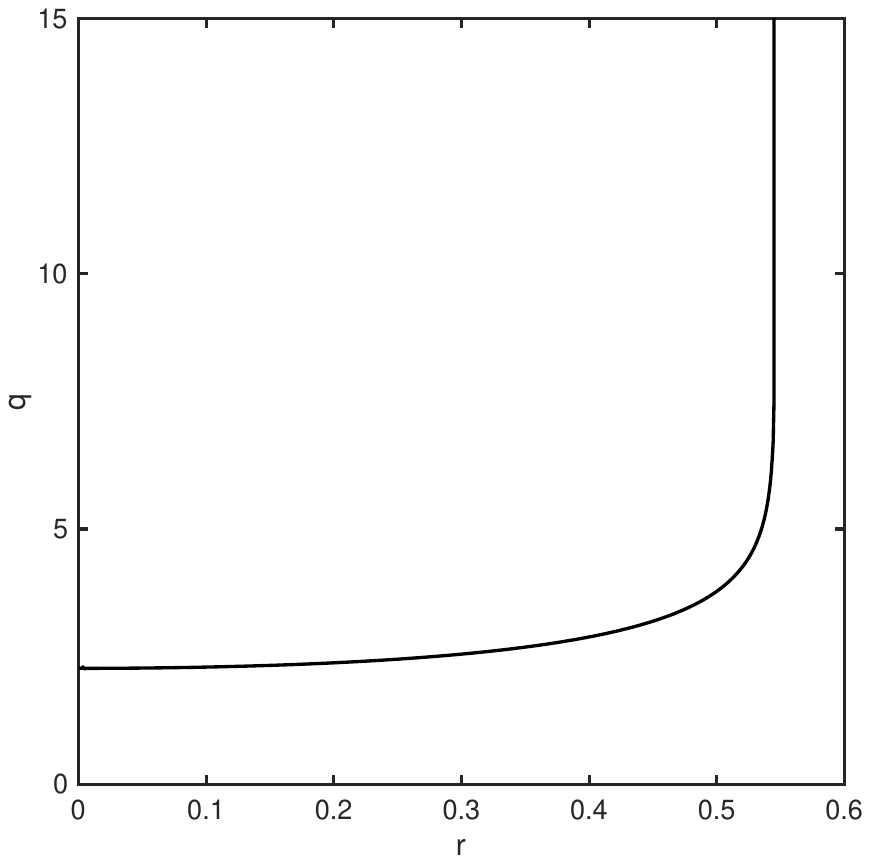}\vspace*{-19mm}
\caption{The surface-averaged safety factor $q$ profiles versus the minor radius on the mid-plane 
   on the low field side for the   
configuration described 
in Fig. \ref{f3}.}
\label{f6}
\end{figure}

\begin{figure}[htp]\vspace*{-21mm}
\centering
\includegraphics[width=80mm]{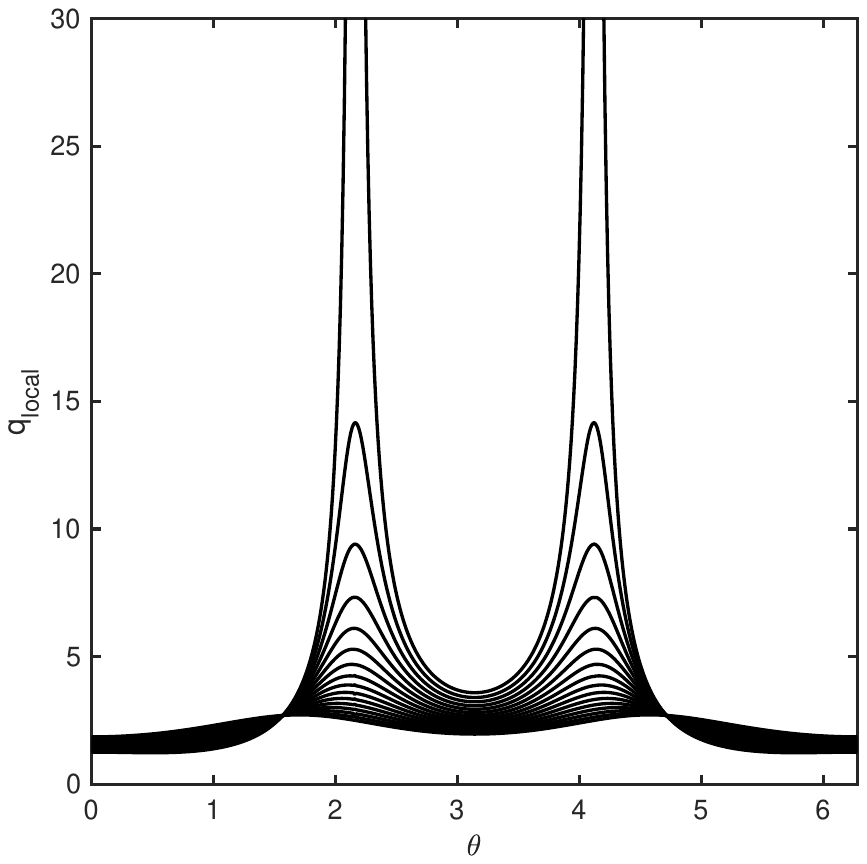}\vspace*{-19mm}
\caption{The local safety factor    ($q_{local}$) profiles versus the poloidal angle for the configuration described 
in Fig. \ref{f3}.   The consecutive   $q_{local}$ are plotted from the plasma center to the edge.}
\label{f7}
\end{figure}

The safety factor can be computed for the equilibria shown in Fig. \ref{f3}.
The results are plotted in Fig. \ref{f6}. 
Because at the X point, $|\bnabla\chi|$ vanishes. The Jacobian in \eq{jac} becomes
infinite. Consequently, as is well known the surface-averaged safety factor becomes infinite
on the last closed flux surface as shown in Fig. \ref{f6}. From the definition of the safety factor in \eq{qq}
one can see that the safety factor is a surface-averaged quantity. 
Noting that the Jacobian only becomes singular at the X points,
one can expect that the integrand in the definition of safety factor, \eq{qq}, is not singular
everywhere. This leads us to plot out the local safety factor profiles, $q_{local}$,
in Fig. \ref{f7}.  From Fig. \ref{f7} one can see that the surface-averaged $q$   alone
as shown in Fig. \ref{f6} may not   completely describe the X-point effects on the MHD modes. The local $q$ depends on the poloidal location.
This leads us to introduce
the     dual-poloidal-region $q$ description in this work in the following sections.
 
   In the current effort, the reasons for using the Solov\'ev equilibrium are as follows.
First, the current work is the first one using the     dual-poloidal-region $q$ description. The analytical
equilibrium allows others in our field to consider this approach and benchmark the results. Second,
for the next efforts to include the nonideal MHD description in $\Theta_X$ and match
the solutions between $\Theta_{core}$ and $\Theta_X$ regions, the analytical equilibrium
simplifies the task a lot.  
The underlying physics of  the mathematical treatment in the dual-poloidal region $q$ description is unchanged with the
Solov\'ev analytical equilibrium.
The method is applicable to the realistic equilibria, which 
is proposed for future studies, perhaps after the non-ideal MHD description of $\Theta_X$ region is developed.
  There seems to be no natural value for choosing $\Theta_X$.  As will be seen in Sec.
\ref{sec.4c},  the choice of $\Theta_X$ bears a similar effect
to where to cut the edge portion in the conventional treatment based on the
surface-averaged $q$ in the ideal MHD description.\cite{trun}

 \section{Dual-poloidal-region safety factor coordinates}

\label{sec.3}

In this section, we describe the flux coordinates with dual-region safety factor in the poloidal direction.
The motivation for this has been discussed in Sections \ref{sec.1} and \ref{sec.2}.
Instead of using the conventional flux coordinates with a surface-averaged safety factor $q$,
the safety factor is defined according to the proximity to the X point in the edge
region of plasma torus, where   
the equilibrium Jacobian becomes large in the local region where $|\bnabla\chi|$
becomes small.

We first describe the dual-poloidal-region $q$ coordinates in the PEST coordinates. The general flux
coordinates will be described afterward. Unlike the conventional flux coordinates in \eq{bb},
the  magnetic field is represented as follows
\bea
\B&=&\bnabla\phi\btimes\bnabla\chi +f(\chi)\bnabla\phi
\nn
&=& \bnabla\phi\btimes\bnabla\chi + \frac{f {\cal J}_{eq}}{X^2} \bnabla\chi\btimes\bnabla\theta_{eq}
\nn
&\equiv&  \bnabla\phi\btimes\bnabla\chi + q_{local}(\chi,\theta_{eq}) \bnabla\chi\btimes\bnabla\theta_{eq},
\label{bq2d}
\eea
  where  $q_{local}(\chi,\theta_{eq}) \equiv f {\cal J}_{eq}/X^2$ represents the local satefty factor.
Here, it is noted that $q_{local}$ depends on the choice of $\theta$.
 To introduce the dual-poloidal-region $q$ 
coordinate system, we note that \eq{bq2d} can be rewritten as
\bea
\B&=& \bnabla\phi\btimes\bnabla\chi + \frac{f \hat {\cal J}_{eq}}{X^2} h(\chi,\theta_{eq})\bnabla\chi\btimes\bnabla\theta_{eq},
\eea
where 
\bean
\hat {\cal J}_{eq}&=& 
\begin{cases}
{\cal J}_{eq}=\frac1{\bnabla\chi\btimes\bnabla\theta_{eq}\bcdot \bnabla\phi},~~~&\theta_{eq}\in \Theta_{core};
\\
 {\cal J}_{eq}^{\max} = const.~~~&\theta_{eq}\in \Theta_X.
\end{cases}
\\
h(\chi,\theta_{eq}) &=& 
\begin{cases}
1,~~~&\theta_{eq}\in \Theta_{core};
\\
\frac{{\cal J}_{eq}}{ J_{eq}^{\max}},~~~&\theta_{eq}\in \Theta_X.
\end{cases}
\eean
Here,  $J_{eq}^{\max}$ is a constant and specifies the maximum allowed value for Jacobian.
   Since the singularity of safety factor results from the Jacobian, we limit its maximum $\jac^{\max}_{eq}$
to pick out the vicinity of the singular point with the boundary of $\Theta_X$  defined by the poloidal angle where
$\jac_{eq} = J_{eq}^{\max}$. Since Jacobian varies both in $\chi$ and $\theta_{eq}$, $\Theta_X$
depends on $\chi$ as well.
Using $\hat {\cal J}_{eq}$ to define the flux coordinates of PEST type, instead of ${\cal J}_{eq}$, one obtains
\bea
\B&=&\bnabla\phi\btimes\bnabla\chi 
+ q_c(\chi) h(\chi,\theta_{c}) \bnabla\chi\btimes\bnabla\theta_{c},
\label{bc0}
\eea
where
\bea
\theta_{c}&=&\frac f {q_c} \int_0^{\theta_{eq}}d\theta_{eq} \frac{\hat J_{eq}}{   X^2 },
\label{thetc}
\\
q_c(\chi)&=&\frac f{2\pi} \int_0^{2\pi}d\theta_{eq} \frac{\hat J_{eq}}{   X^2}.
\label{qc}
\eea
Here,  we have used \eq{thetc} for $\theta_c(\theta_{eq})$ 
to transform $h(\chi,\theta_{eq})$ to $h(\chi,\theta_{c})$.
  Equation \eqn{bc0} expresses the magnetic field in the dual-poloidal-region $q$ coordinates
$(\chi,\theta_c,\phi)$.
In the poloidally core region $\Theta_{core}$,  the coordinate system $(\chi,\theta_c,\phi)$ becomes the 
straight-field-line flux coordinates with the finite safety factor $q_c(\chi)$ 
being a function of $\chi$ only as in \eq{bb}. In the vicinity
of X point $\Theta_X$, the local safety factor becomes $q_{local} = q_c(\chi) h(\chi,\theta_c)$,
which is a function of $\chi$ and $\theta_c$.  Therefore, the dual-poloidal-region $q$ coordinates
are different from the conventional single $q$ coordinates given in \eq{bb}  but are
one of the general coordinate system given in \eq{bq2d}.
  Note here that since the Jacobian, $\hat\jac_{eq}$,  is defined according to    the given 
  $\jac^{\max}_{eq}$,
which determines  $\Theta_X$, 
the parameters $\theta_c$, $q_c$, and $h$ in the dual-poloidal-region $q$
coordinates depend on   $\jac^{\max}_{eq}$ or $\Theta_X$.

Now, let us describe the conversion to the general flux coordinates of dual-poloidal-region $q$.
Letting
\bean
\zeta_{gc} = \phi +  \nu_{eq} (\chi,\theta_c)~~~\hbox{and}~~~ \theta_{gc} 
=\theta_c+  \nu_{eq}(\chi,\theta_c) /(q_c h),
\eean
one has
\bean
&&\bnabla\phi\btimes\bnabla\chi  =
 \bnabla\zeta_{gc}\btimes\bnabla\chi - \pd   \nu_{eq},{\theta_c} \bnabla\theta_c\btimes\bnabla\chi,  
\\
&&q_c h(\chi,\theta_{c}) \bnabla\chi\btimes\bnabla\theta_{c}
=q_c h\bnabla\chi\btimes\bnabla\theta_{gc} 
\nn&&
 ~~~~~ - \pd   \nu_{eq},{\theta_c} \bnabla\chi\btimes\bnabla\theta_{c},
  -  \nu_{eq} h \pd,{\theta_{gc}}\lbs \frac 1 {h}\rbs \bnabla\chi\btimes\bnabla\theta_{gc}.
\eean
 
Here, $  \nu_{eq}$ is an arbitrary periodic function of $\theta_c$. One can specify it according
to the specific coordinates to be chosen, such as the Hamada or Boozer coordinates.\cite{zb1}
Therefore, the magnetic field in \eq{bc0} can be transformed to
\bean
\B&=& \bnabla\zeta_{cg}\btimes\bnabla\chi 
+ q_c h_g(\chi,\theta_{gc}) \bnabla\chi\btimes\bnabla\theta_{gc},
\eean
where
\bean
h_g(\chi,\theta_{gc}) &=& 
\begin{cases}
1,~~~\theta_{gc}\in \Theta_{core};
\\
\frac{J_{eq}}{{ \cal J}_{eq}^{\max}}
\lbm 1
+ \frac{  \nu_{eq}}{q_c} \pd,{\theta_{gc}}\lbs \frac{{\cal J}_{eq}^{\max}}{ {\cal J}_{eq}}\rbs \rbm,~\theta_{gc}\in \Theta_X .
\end{cases}
\eean

Next, let us discuss why we introduce the dual-poloidal-region $q$ coordinates.
From the definition of surface-averaged $q$ in \eq{qq} one can see that the integrand ($\jac$) becomes
infinite only at X points. It is the average that makes $q$ become infinite everywhere on the last closed flux surface.
Mathematically, we know that the singular point should be isolated with a subtle treatment,
instead of being  averaged to spread it.   In our field, we often see  this type of treatments,
for example, the singular layer theory for treating the resonance surfaces.\cite{mer,gj,dcon} 
The dual-poloidal-region $q$ coordinate treatment is an effort in this direction.

Note that if one uses the magnetic field representation
with the surface-averaged $q$ as given in \eq{bb},
the MHD energy needs to be minimized individually in
each radial interval  between  two neighboring resonance surfaces. Since there are infinite number
of the intervals using the surface-averaged $q$, this makes the conventional treatment practically 
 inapplicable for the equilibria with X points. 

In the poloidal direction,  the single $q$ description leads the main harmonic to be $m^{single~q}_{main} = nq$, which is infinite, using the Fourier decomposition method.  The usual numerical scheme  is aimed at  
minimizing the shear Alfv\'en energy of main harmonic 
 for physics investigation. Being unable to treat the case with 
$m^{single~q}_{main} = nq\to \infty$  
just lets it to miss the physics goal for X point physics. 
Instead, in our dual $q$ description the main harmonic becomes  $m^{dual~q}_{main} = nq_c$, which is finite. 
The finite $m^{dual~q}_{main}$ exactly reflects the filament feature observed experimentally  in Fig. 1,
which are aligned to the local magnetic field. 

Reference \onlinecite{dcon} gives the physics mechanism of energy minimization
 process, which other codes cannot avoid either as soon as the surface-averaged $q$ alone is used. 
 To minimize the  shear Alfv\'en energy on a surface with infinite $q$, the grids have to be infinitely fine.  
Therefore, this issue is beyond the Fourier decomposition method.
  We particularly note that in the finite-element-based codes
 sometimes the safety factor is used as the weighting
factor in the assignment of the mesh points around the resonance surfaces.\cite{gato} 
 In the the dual-poloidal-region $q$ coordinates, however,
 the singularity is forced to retreat back to the vicinity
of X point, $\Theta_X$, as it originally locates. This helps to solve the  difficulty.

Besides, the surface-averaged $q$ not only leads $q$ become infinite as approaching to the last closed
flux surface but also the magnetic shear. 
We know that when the magnetic shear is big, the non-ideal MHD
effects, such as the FLR and resistivity effects, need to be included.
This causes that even the poloidally core region, $\Theta_{core}$, needs formally a nonideal MHD treatment
in the single $q$ description since
the distance between the neighboring resonance surfaces become smaller than the Larmor radius.
In fact, the local $q$ and magnetic shear are finite in the poloidally core region $\Theta_{core}$. 
The dual-poloidal-region $q$ coordinates help to solve the difficulty and open the path to treat the
poloidally core region $\Theta_{core}$ as ideal MHD, but the vicinity of X points $\Theta_X$ as 
nonideal MHD, and then match them together.  
This is another distinct feature of the dual-poloidal-region $q$ description.

Furthermore, we point out that 
it has been realized before that the local safety factor diverges only in the vicinity of X points.
Various MHD codes have been developed aiming at taking into consideration this specialty,
for example GATO and KINX.\cite{gato,kinx} Often the finite element method is used. 
Our approach is one of these efforts. 
Note that in the vicinity of X points, both the local safety factor and magnetic shear
tend to infinity. In the poloidal direction, the distance between the same magnetic field lines
in the different toroidal loops, $\delta L_X$, tends to zero. In the radial direction, a big magnetic shear
appears. They makes the numeral approach have to have infinitely fine grids
to resolve in the ideal MHD description. 
  Note that the singular layer theory in which Mercier's criterion is derived is in principle unaffected (at least asymptotically) by the presence of X points.\cite{mer,gj,lor,has}
 This is because theoretically one can assume
the modes to be infinitely localized.
However, this type of minimizations cannot be achieved 
numerically with finite grid density.
The singular boundary layer theory in the poloidal direction for $\Theta_X$ is needed or the nonideal MHD physics 
needs to be added in this case.  Therefore, formally isolating the $\Theta_X$ region
in the dual-poloidal-region coordinates  helps. 
The  $\Theta_X$ region
eventually needs the nonideal MHD description.   Further discussion of 
dual-poloidal-region $q$ coordinates can be found in Appendix \ref{app.cood}.

These complete the description of the dual-poloidal-region $q$ coordinates. They will be used to study the edge
localized modes in the following sections.

 \section{X-point effects on   the ideal MHD modes}

\label{sec.4}

In this section, we study the X-point effects on    
the ideal MHD modes in tokamaks with the coordinates of 
dual-poloidal-region safety factor developed in Sec. \ref{sec.3}. 
  Since the X-point effects mainly affect the edge region, 
the modes localized at tokamak edge are particularly examined.
The magnetic field in this coordinate system can be expressed as follows
\bea
\B&=& \bnabla\phi\btimes\bnabla\chi 
+ q_c(\chi) h(\chi,\theta_c)\bnabla\chi\btimes\bnabla\theta_{c}
\label{bt0}
\\
&=& \B_c +\B_X,
\label{bt}
\eea
where
\bea
\B_c&=&\bnabla\phi\btimes\bnabla\chi 
+ q_c(\chi) \bnabla\chi\btimes\bnabla\theta_{c},
\label{bc}
\\
\B_X&=&
\begin{cases}
0,&\theta_{gc}\in \Theta_{core};
\\
q_c(\chi)(h-1) \bnabla\chi\btimes\bnabla\theta_{c},&\theta_{gc}\in \Theta_X .
\end{cases}
\label{bx}
\eea
The coordinates set apart the region where the safety factor is very large, $\Theta_X$. 
In the region $\Theta_X$, the field lines are almost in the toroidal direction, i.e, 
\bean
\B &\approx& I\bnabla \phi,~~~ \theta \in \Theta_X.
\eean

As discussed earlier, we do not minimize the energy with the surface-averaged $q$ coordinates. The resonance condition
$m-nq=0$ in    the Fourier decomposition 
treatment can cause the poloidal mode number m to become infinite. 
To reflect that the filaments are aligned with the local magnetic field line as shown in the experiments, 
in this section we first consider the peeling modes
in the field $\B_c$, while taking into account the $\B_X$ effects in the region $\Theta_X$.  Considering
$\B_c$ alone in the peeling or peeling-ballooning studies resembles the conventional numerical
treatment by truncating the tokamak edge portion.\cite{trun}  As will be seen, including the 
additional effects from $\B_X$ in the
region $\Theta_X$ gives rise to the X-point stabilization effect.
In the vicinity of X points, the toroidal field becomes dominant, i.e., the local $q$ is infinite. One can imagine
that the poloidally localized asymmetric modes can develop there, which will be addressed next.
At the end of this section, the numerical calculation based on the code AEGIS-X is presented with the
full magnetic field $\B$ in \eq{bt0} taken into account,\cite{aegis} which confirms  the existence of
MHD filaments aligned with the local field line in the region $\Theta_X$ and the stabilization effects
of  X points.

 \subsection{The peeling type of modes}

\label{sec.4a}

In this subsection, we first study the X-point effects on  the peeling type of modes. 
To reflect the experimental observation that the filament is aligned with the local field line,
to imitate the conventional treatment by truncating the edge portion, and also to avoid
the modes resonating at infinite m due to using the surface-averaged $q$,
we will minimize the energy principle with the magnetic field $\B_c$ in \eq{bc}
in the whole region: $\Theta_{core}$ and $\Theta_X$ and then add the modification by $\B_X$ 
in \eq{bx} in the vicinity of X points, $\Theta_X$,  afterward. In this way, the X-point effects 
are explained.

 The energy principle   is used for this study \cite{gre}
\bea
2\delta W 
&=& \int \bigg\{ \frac1{\mu_0}\left |\delta \B -\B\frac{\mu_0\zxi\bcdot\bnabla P}{B^2} \right |^2
- \frac{j_\|}B  \zxi\btimes\B\bcdot\delta\B 
\nn&&
- 2(\zxi\bcdot \bnabla P)(\zxi \bcdot \Kb) 
+\Gamma P( \bnabla \bcdot \zxi)^2 
\bigg\} d\r,
\label{ene}
\eea
where $\zxi$ denotes the field line displacement, which is related to the perturbed magnetic
field $\delta \B =\bnabla \btimes\zxi\btimes\B$, 
$\j$  is the equilibrium current density, 
$\mu_0\delta \j =\bnabla\btimes 
\delta\B$ denotes the perturbed current density,   $\Kb$ is the field line curvature,
the subscripts $\bot$ and $\|$ denote the perpendicular and parallel to the equilibrium magnetic field,
 $\mu_0$ is magnetic constant, 
 the perturbed pressure  $\delta P = -\zxi\bcdot \bnabla P -\Gamma \bnabla \bcdot \zxi$, 
  $\Gamma$ denotes the ratio of specific heats, and vectors are denoted by boldface.

We first consider energy minimization under the equilibrium determined by
$\B_c$ in \eq{bc}. In this case, $\j_c =\bnabla \btimes\B_c$ and $\bnabla P_c = \j_c\btimes\B_c$.
The minimization process is similar to the derivation of the Mercier criterion and the
stability criterion for peeling modes.\cite{mer,gj,lor,has}
For brevity, the subscript c is omitted.
Following Ref. \onlinecite{gre}, the Hamada coordinates are used with
$\psi$ and $\chi$ denoting the toroidal and poloidal magnetic fluxes, 
$Z$ labeling the magnetic surfaces, and $Z_0$ being the reference resonance surface.
We introduce the localized flux coordinates:
\bea
x&=&Z-Z_0,
\label{4ax}
\\
u&=&\psi'(Z_0)\theta-\chi'(Z_0)\zeta.
\label{4au}
\eea
Therefore, the parallel derivative becomes
\bean
\B\bcdot\bnabla &=&\frac{\psi'}{V'} \pd ,\zeta+\frac{\Lambda x}{V'}\pd, u
\eean
with $V$ being the volume inside a flux surface, prime denoting the derivative with respect to $Z$, and
\bean
\Lambda &=& \psi'(Z_0)\chi''(z_0) -\chi'(Z_0)\psi''(z_0).
\eean
It is also defined 
\bean
\Theta&=&\frac{\bnabla Z\bcdot\bnabla u}{|\bnabla Z|^2}
\eean

We  decompose the field line displacement as
\bea
\zxi&=& \xi\frac{\bnabla Z}{|\bnabla Z|^2} +\mu\frac{\B_c\btimes \bnabla Z}{B_c^2} +\nu\frac{\B_c}{B_c^2}
\label{xi}
\eea
and introduce the following orderings:  \cite{glasser}
\bean
x\sim \epsilon\ll 1,~~~~ \pd ,V\sim\epsilon^{-1},~~~ \pd ,u\sim\pd,\zeta \sim1.
\eean
Therefore, we can further assume the following ordering scheme, which can be proved a posteriori 
\bea
\xi  =\epsilon \xi^{(1)}+\cdots,~~~\mu=\mu^{(0)} +\cdots,~~~\nu=\nu^{(1)}+\cdots.
\label{ord1}
\eea
Here, the superscripts are used to indicate the orderings.

The minimization process for localized modes is standard.\cite{mer,gj,lor,has} One can find detailed derivation in
Ref. \onlinecite{zb1}. 
 One finally obtains
\bea
\delta W_c &=&\frac12 M\omega^2 \int  |\xi|^2dx
\nn
&=&  \frac{c_0}2\lbl \int\lbm x^2\lbs \od\xi,x\rbs^2   - \lbs  D_I+\frac14 \rbs \xi^2 \rbm dx
\right.
\nn&&\left.
+ \left.\lbs \Delta +\frac12\rbs x\xi^2\right |_a^b\rbl,
 \label{ene1}
\eea
where
\bean
c_0&=&\frac{\Lambda^2}{\mu_0 }\frac{\lbm\oint (dl/B)\rbm^2}{\oint (B^2/|\bnabla V|^2)(dl/B)},
\nn
\Delta &\equiv&\frac12 -\frac1{\Lambda } \av{\frac{\sigma B^2}{|\bnabla Z|^2}},
\nn
D_I&\equiv& E+F+H-\frac14,
\nn
E&\equiv&\frac{\av{B^2/|\nabla V|^2}}{\Lambda^2}
\lbs J'\psi''-I'\chi''+
\Lambda\frac{\av{\sigma B^2}}{\av{B^2}}\rbs,
\nn
F&\equiv&\frac{\av{B^2/|\nabla V|^2}}{\Lambda^2}
\lbs\av{\frac{\sigma^2 B^2}{|\nabla V|^2}}-
\frac{\av{\sigma B^2/|\nabla V|^2}^2}
{\av{B^2/|\nabla V|^2}}
\right.\nn&&\left.
+P'^2\av{\frac1{B^2}}\rbs,
\nn
H&\equiv&
\frac{\av{B^2/|\nabla V|^2}}{\Lambda}
\lbs
\frac{\av{\sigma B^2/|\nabla V|^2}}
{\av{B^2/|\nabla V|^2}}-\frac{\av{\sigma B^2}}{\av{B^2}}
\rbs,
\eean
  $\sigma=\J\bcdot\B/B^2$, and $M= M_c+M_t$ with
\bean
M_c&\equiv&
\frac{\rho_m}{\alpha^2\Lambda^2}\av{\frac{B^2}{|\nabla V|^2}}
\av{\frac{|\nabla V|^2}{B^2}}
,
\label{I2.mc}
\\
M_t&\equiv&
\frac{\rho_m}{\alpha^2\Lambda^2{P'}^2}
\av{\frac{B^2}{|\nabla V|^2}}
\lbs\av{\sigma^2B^2}-
\frac{\av{\sigma B^2}^2}{\av{B^2}}\rbs.
\label{I2.mt}
\eean
  Here, it has been noted that the vacuum energy can be neglected for peeling modes
as proved in Ref. \onlinecite{lor}.

After minimization of the total energy in \eq{ene1} with respect to $\xi$,
one obtains the singular layer equation
\bea
\od ,x  x^2  \od \xi,x - \lbs \frac14 + D_I\rbs \xi =0.
\label{sin}
\eea
Its solution is
\bea
\xi=\xi_0 |x|^{-\frac12\pm \sqrt{-D_I} }.
\label{peelsol}
\eea
Inserting \eq{sin} into \eq{ene1},  the energy principle is reduced to
\bea
\delta W_c&\ge& c_0
\lbm \frac {x^2}2 \lbs \xi^*\od \xi,x +\xi\od \xi^*,x\rbs + \lbs \Delta +\frac12\rbs x |\xi|^2\rbm^{x_b}_{x_a}.
\label{peel0}
\eea
Then, by inserting the solution of the singular layer equation in \eq{peelsol}, one can find the minimum energy. 
Without considering the X-point contribution, the stability condition for peeling mode is just $\Delta<0$.
One needs to add the X-point effects from $\B_X$ in \eq{bx} in the vicinity of X point, $\Theta_X$.

To find out the $\B_X$ effects seems to be complicated.
However, we note that 
the region $\Theta_X$ is smaller than $\Theta_{core}+\Theta_X$ and more importantly that
the minimization process for the localized modes leading to \eq{ene1} has already been performed
in the magnetic field described by $\B_c$. Therefore, the non-minimized  
Alfv\'en mode contribution in the first term
of \eq{ene} is dominant and we only need to add the additional contribution to this term in region $\Theta_X$.

First, we note that $\delta \B$ in  \eq{ene}  can be decomposed as 
\bea
&&\delta \B \bcdot \bnabla Z = \bnabla\bcdot \lbm\lbs \zxi\btimes \B\rbs \btimes \bnabla Z \rbm
\nn&&\hspace*{15mm}
=\B\bcdot\bnabla (\zxi\bcdot\bnabla Z),
\label{dbr}
\\
&&\frac{\delta \B \bcdot \B\btimes\bnabla Z }{|\bnabla Z|^2} 
= \B\bcdot \bnabla \lbs \frac{\zxi\bcdot \B\btimes\bnabla Z}{|\bnabla Z|^2}\rbs 
\nn
&&\hspace*{15mm}
-\frac{\B\btimes\bnabla Z}{|\bnabla Z|^2} \bcdot  \bnabla\btimes \frac{\B\btimes\bnabla Z}{|\bnabla Z|^2}\zxi\bcdot\bnabla Z,
\label{dbt}
\\
&&\frac1{B^2} \lbs \delta\B -\B \frac{\mu_0\zxi\bcdot\bnabla P}{B^2}\rbs\bcdot \B=- \bnabla\bcdot\zxi
\nn&&\hspace*{15mm}
 -\B\bcdot\bnabla\lbs\frac{\zxi\bcdot\B}{B^2}\rbs-
2\zxi\bcdot\Kb.
\label{dbpar}
\eea
Here, we have considered the decompositions in the total magnetic field in \eq{bt}, instead of $\B_c$, in
order to take into account the extra energy from the X-point contribution.

To evaluate the $\delta \B$ components in Eqs. \eqn{dbr}-\eqn{dbpar}, we decompose the field
line displacement in the minimization process with $\B_c$ in \eq{xi} in the total magnetic field representation
\bea
\zxi&=& \xi_t\frac{\bnabla Z}{|\bnabla Z|^2} +\mu_t\frac{\B\btimes \bnabla Z}{B^2} +\nu_t\frac{\B}{B^2}.
\label{xix}
\eea
Equating Eqs. \eqn{xi} and \eqn{xix}  one obtains
\bea
\xi_t&=& \xi,
\label{xit}
\\
\mu_t&=&\mu \frac{\B\bcdot\B_c}{B_c^2},
\label{mut}
\\
\nu_t&=& \nu \frac{\B\bcdot\B_c}{B_c^2}.
\label{nut}
\eea
Here, it has been noted that the poloidal magnetic field is negligible in the region $\Theta_X$
and $\B\bcdot \B_c= f q_c/{\cal J}_c$. 

Using Eqs. \eqn{dbr}-\eqn{nut}, one can evaluate the extra energy from the X-point effects by
calculating the first term of \eq{ene}.   This is because the first term represents
the energy of the Alfv\'en modes, which is one order larger than the rest terms in the
singular layer theory.\cite{mer,gj} 
Noting that in the minimization process with $\B_c$, one has
$\mu$ is one order larger than $\xi$ and $\nu$.\cite{gre,zb1}  Therefore, the extra energy
from the X-point contribution becomes
\bea
\delta W_X 
&=& \frac1{2\mu_0}\int_{\Theta_X} \left |\delta \B -\B\frac{\mu_0\zxi\bcdot\bnabla P}{B^2} \right |^2 d\r
\nn
&=& \frac1{2\mu_0}\int_{\Theta_X}  \left | \frac{|\bnabla Z|}B \B\bcdot \bnabla \lbs \frac{\zxi\bcdot \B\btimes\bnabla Z}{|\bnabla Z|^2}\rbs \right |^2 d\r 
\nn
&=& \frac1{2\mu_0}\int_{\Theta_X}  \left | \frac{|\bnabla Z|}B \B\bcdot \bnabla \lbs
\mu_t \frac{B^2}{|\bnabla Z|^2}\rbs \right |^2 d\r 
\nn
&=& \frac{n^2}{2\mu_0 }\int_{\Theta_X}  \left | \frac {B \B\bcdot\B_c} {B_c^2|\bnabla Z|}  \frac{f}{X} \mu \right|^2d\r. 
\label{enex}
\eea
Here, $\mu$ is related $\xi$ in $\delta W_c$ in \eq{peel0} by\cite{gre} 
\bean
\pd \xi,x+\pd \mu,u=0.
\eean
From the ordering analyses in \eq{ord1}, one can see that the integrand in $\delta W_X$
in \eq{enex} is larger than that in $\delta W_c$ in \eq{peel0}. However,
the integration domain for $\delta W_X$, $\Theta_X$, is smaller than that for $\delta W_c$, $\Theta_{core}+\Theta_X$.

The system stability is determined by the total energy
\bean
\delta W =\delta W_c+\delta W_X.
\eean
Since $\delta W_X$ is positive definite, the X-point effects are shown to give rise to the stabilizing 
effects. 
The    smaller the region $\Theta_X$, the more   stringent the stability condition becomes.  Only when  $\B$ becomes
parallel to $\B_c$ in the limit $\Theta_X\to0$,  
one has $\delta W_X =0$.
This  indicates that the usual numerical treatment by truncating the tokamak edge region
gives rise to an overly stringent stability condition. 
 Smaller $\Theta_x$ implies the larger $q_c$. As discussed earlier   in Sec. \ref{sec.3}, 
there is a physics   and numerical  limitation for approaching the surface-averaged safety factor. 
These phenomena will be further confirmed numerically using AEGIS-X 
code to be described in Sec. \ref{sec.4c}.

\subsection{The axisymmetric modes in the vicinity of X points}

\label{sec.4b}

In this subsection, we discuss a special mode  
which can be recovered in
the dual-poloidal-region $q$ description, the axisymmetric modes in the vicinity of X points.
In the surface-averaged $q$ description, the conventional safety factor becomes infinite
on the last closed flux surface. However, it does not imply the poloidal magnetic field vanishes
everywhere on the last closed flux surface. Actually, the local $q$ in the poloidally core region, $\Theta_{core}$
remains finite. The magnetic field becomes basically the toroidal field
only in the vicinity of X points in the region $\Theta_X$. Because
the magnetic field is about along the toroidal direction in $\Theta_X$, 
  in view of the minimization of the field line bending effects
one may expect that the $n=0$ localized axisymmetric modes can develop in this region.
  Mathematically, this situation is well described by the
dual $q$ coordinates. From the description of the dual-poloidal-region $q$ coordinates in Sec. \ref{sec.3}
one can see that $h=1$ in $\Theta_{core}$ and $h=\jac_{eq}/\jac_{eq}^{\max}$ in $\Theta_X$. 
This makes $q_ch$ become $q_{local}$ in the vicinity of X points, which is infinite. 
Therefore, one can carry directly the axisymmetric mode analyses in $\Theta_X$
in the dual-poloidal-region $q$ description.

 The localized stability criterion for $n=0$ axisymmetric modes
  can be obtained from the conventional Mercier or peeling stability criterion, which
 is actually embedded in the discussion of peeling modes in Sec. \ref{sec.4a}, 
 \bea
\delta W_c&=& \frac{ c_0}2\lbl \int_{  \Delta\Theta_X}\lbm x^2\lbs \od\xi,x\rbs^2   - \lbs  D_I+\frac14 \rbs \xi^2 \rbm dx
\right.\nn&&\left.
+\left.\lbs \Delta +\frac12\rbs x\xi^2\right |_a^b\rbl,
 \label{ene2}
\eea
where   $\Delta\Theta_X$ is the region in the vicinity of X point where the
the axisymmetric perturbation is considered,
\bea
c_0&=&\frac{\Lambda^2}{\mu_0 }\frac{\lbm\oint (dl/B)\rbm^2}{\oint (B^2/|\bnabla V|^2)(dl/B)}
\nn
&\approx& \frac{2\pi X_X \Lambda^2}{\mu_0 }\frac{|\nabla V|^2}{B^3},
\\
\Delta &\equiv&\frac12 -\frac1{\Lambda } \av{\frac{\sigma B^2}{|\bnabla Z|^2}}
\approx \frac12 -\frac1{\Lambda } \frac{\sigma B^2}{|\bnabla Z|^2},
\\
D_I&\equiv& E+F+H-\frac14,
\label{di}
\\
E&\equiv&\frac{\av{B^2/|\nabla V|^2}}{\Lambda^2}
\lbs J'\psi''-I'\chi''+
\Lambda\frac{\av{\sigma B^2}}{\av{B^2}}\rbs
\nn
&\approx&\frac{B^2/|\nabla V|^2}{\Lambda^2}
\lbs J'\psi''-I'\chi''+
\Lambda\sigma \rbs,
\\
F&\equiv&\frac{\av{B^2/|\nabla V|^2}}{\Lambda^2}
\lbs\av{\frac{\sigma^2 B^2}{|\nabla V|^2}}-
\frac{\av{\sigma B^2/|\nabla V|^2}^2}
{\av{B^2/|\nabla V|^2}}
\right.\nn&&\left.
+P'^2\av{\frac1{B^2}}\rbs
\approx\frac{P'^2}{\Lambda^2/|\nabla V|^2},
\\
H&\equiv&
\frac{\av{B^2/|\nabla V|^2}}{\Lambda}
\lbs
\frac{\av{\sigma B^2/|\nabla V|^2}}
{\av{B^2/|\nabla V|^2}}-\frac{\av{\sigma B^2}}{\av{B^2}} 
\rbs
\nn
& \approx& 0
\label{hhh}
\eea
with $X_X$ being the major radius of X point. 
Here, it has been considered that the equilibrium is axisymmetric, the poloidal magnetic field
about vanishes, and the poloidally localized $n=0$ modes are considered.   
  In this case, the average $\av{\cdot}$ here represents the avera
  ge over a toroidal loop.
The Pfirsch-Schl\"uter current is negligible 
and the parallel current density $\sigma$  is dominated 
by the axisymmetric Ohmic current in the vicinity of X points.

Like the Mercier criterion, the stability condition  just indicates the condition of mode existence. 
  Since $q_{local}\to \infty$, the field lines   in the region $\Theta_X$  just run toroidally 
 In this case,
the stability criterion reduces to the conventional flux-tube-like analyses for
interchange modes with magnetic shear taken into account. 
 Besides the physical interpretation of $E$, $F$, and $H$ representations in Ref. \onlinecite{glasser}
and the expressions in a tokamak with a  large aspect ratio and circular cross section in Ref. \onlinecite{glasser1},
 we use an alternative approach to show
the existence condition of the axisymmetric  modes.
Note that \eq{di} can be alternatively written as\cite{bal1,bal2}  
 \bea
D_I&=&\frac{\av{g}}{\av{\B\cdot\nabla\Lambda_s}^2}
\lbm\av{P'\kappa_n}+\av{\B\cdot\nabla\Lambda_s
\lbs\lambda_c-\frac{\av{g\lambda_c}}{\av{ g}}\rbs}
\right.
\nn
&&\left.
+\av{g\lbs\lambda_c^2-\frac{\av{g\lambda_c}^2}{\av{ g}^2}\rbs}\rbm-\frac14,
\label{di3}
\eea
where
\bean
&&g=\frac {B^2}{|\bnabla \chi|^2},~~\kappa_n=
\frac{2\nabla \chi\cdot\Kb}{|\nabla \chi|^2},
\nn&&
\kappa_g=
\frac{\B\times\nabla \chi\cdot\Kb}{B^2}, 
~~~
\B\bcdot \bnabla \lambda_c=\mu_0P'\kappa_g,
\nn
&&
\B\cdot\nabla\Lambda_s=-\frac 1{|\nabla\chi|^4}
(\B\times\nabla\chi)\cdot\nabla \times(\B\times\nabla\chi).
\eean
Here, we have kept the notations in Ref. \onlinecite{bal2}. The formulas to prove the equivalence 
between Eqs. \eqn{di} and \eqn{di3} can be found in Ref. \onlinecite{zb1} with $\lambda_c=\sigma/2$.
In the localized axisymmetric mode limit in the vicinity of X point,  \eq{di3} is reduced to
 \bea
D_I&\approx&\frac{g\mu_0P'\kappa_n}{(\B\cdot\nabla\Lambda_s)^2}   -\frac14.
\label{di4}
\eea
Note further that the magnetic shear parameter can be reduced to 
\bean
&&\B\cdot\nabla\Lambda_s
\nn
&=&-\frac 1{|\nabla\chi|^4}
(\B\times\nabla\chi)\cdot\nabla \times\lbm (q_c h \bnabla \chi \btimes\bnabla\theta_c) \times\nabla\chi\rbm 
\nn
&\approx&\frac 1{|\nabla\chi|^4} \bnabla (q_ch)\btimes
(\B\times\nabla\chi)\cdot \lbm (\bnabla \chi \btimes\bnabla\theta_c) \times\nabla\chi\rbm 
\nn
&=&\frac{\B\bcdot\bnabla\theta_c}{|\bnabla\chi|^2} 
\nabla\chi \bcdot\bnabla (q_ch).
\eean
The physical meaning is obvious here. The term $gP'\kappa_n$ denotes the well-known 
magnetic well effect. The term $1/4$ is related to the magnetic shear $(\B\cdot\nabla\Lambda_s)^2$
as relatively compared with the first term on the right-hand side of \eq{di4}.

In the ordering analyses in \eq{est1} in Appendix \ref{sec.appendix},  we show that the first term
on the right-hand side of \eq{di4} becomes
\bea
\frac{g\mu_0P'\kappa_n}{(\B\cdot\nabla\Lambda_s)^2} \sim  (X_s^2 \mu_0/B^2) \Kb\bcdot\bnabla p.
\label{estn}
\eea
The stability condition in \eq{di4} indicates that the estimate in  \eq{estn} should be larger that
$1/4$ for the instability to occur  in the ideal MHD description. 
It may happen. However, the condition is high especially as compared with the resistive MHD case.
When the resistivity is taken into 
account, one can expect that 
the $n = 0$ resistive interchange and tearing modes can develop 
  on the bad curvature side in the vicinity of X point.\cite{glasser} 
  As proved by the well-known theory in Ref. \onlinecite{glasser}, 
when the resistivity effects are taken into account, the magnetic shear stabilization term, i.e., the
term ``1/4"   (which is related to $(\B\bcdot\bnabla\Lambda)^2$), in \eq{di4} disappears. This causes the resistive modes to develop   as soon as $\Kb\bcdot\bnabla p>0$ in the vicinity of X points. 

The localized axisymmetric modes can be understood by comparing a tokamak scenario
without a toroidal current to generate the poloidal magnetic field for field line rotational transform
$1/q$, which  is unstable to the interchange modes on the low field side in this case.
Tokamak stability relies on the average magnetic well, which is induced
by the finite safety factor. 
The difference for the  localized axisymmetric modes 
is  that there is a large magnetic shear in the vicinity of X points. 
However, the resistivity can delete the shear stabilization effects.\cite{glasser} 
Therefore, the localized axisymmetric modes can potentially develop 
in the vicinity of X points.

Furthermore, noting the axisymmetric mode feature, one can potentially    apply
 the  $n=0$   or high $m/n$
  resonance magnetic perturbation in the region near the X point to enhance the X-point
 transport to mitigate ELMs. 
 Note that we are discussing the axisymmetric RMP,  
an external drive here. It does not necessarily require the high $m$ axisymmetric modes to be unstable.
It can happen when the magnetic field patterns coincide.
In fact, neither the conventional $n\not=0$ RMP is considered to resonate 
with the peeling-ballooning  modes
at the pedestal. Instead, the finite $m/n$ magnetic field pattern resonance is considered.
 Engineeringly, this   can be an alternative  the current finite
 $m/n$ RMP approach.  One can also expect that the current dual-poloidal-region $q$ description
 can affect the understanding of the cross field line transport in the divertor region.\cite{evan,rmp1}

\subsection{X-point effects on the   external kink  modes}

\label{sec.4c}

In the subsections \ref{sec.4a} and \ref{sec.4b},
we have demonstrated the X-point effects on the localized modes analytically.
In this subsection, we show the numerical results about the existence of the modes
aligned with the local magnetic field lines in the poloidally core region, $\Theta_{core}$, and
 the X-point effects on them. 
 The numerical treatment allows us to include
the coupling of multiple Fourier components to actually deal with
the   external kink   modes. 
It is noted that in the later peeling-ballooning mode computations, the external
kink modes are sometimes grouped into the low $n$ peeling-ballooning modes.
This is an extension of the analytical theory 
for peeling modes in Sec. \ref{sec.4a}. The axisymmetric modes described in
Sec. \ref{sec.4b} are trivial to understand and therefore are not specially 
needed to have a numerical demonstration
in the current status.

We extend  the AEGIS code\cite{aegis}  to AEGIS-X for this study by numerically solving the MHD equation as follows
\bean
-\rho_m\omega(\omega-\omega_{*})\zxi_\bot&=& \delta \J\btimes \B +\J\btimes
\delta \B 
-\bnabla \delta P.
\eean
Here, for simplicity we consider only imcompressible plasma, i.e., ignoring the coupling of 
parallel motion. One reason for this also is because the connection length between the good and
bad curvature regions varies significantly from the poloidally core ($\Theta_{core}$) to X point ($\Theta_X$) regions.
More subtle kinetic treatment for parallel motion is required in this case. This is beyond 
the current MHD framework. We do include  the ion diamagnetic drift effect, $\omega_*$,
in the inertia term. This can suppress the high m harmonic coupling. 

In AEGIS, \cite{aegis} the Fourier decomposition method
is used in the poloidal direction,   which precludes numerically including
the diverted surface like other codes based on the Fourier decomposition method.
 In the radial direction, the decomposition based on the
independent solution method is used. The adaptive shooting method is then
used to obtain the independent solutions. 
The conventional general flux coordinates with the surface-averaged $q$ are used in the
AEGIS code. In the presence of X points, the safety factor tends to infinity on the last closed flux surface.
One may be able to minimize the field line bending effects in the general straight field line coordinates.
However, the minimization of resonance effects, $m-nq=0$, leads the poloidal mode number to become infinite. 
This is certainly unacceptable   in numerical treatment    for MHD description.
Noting that infinite   local $q$ actually occurs only in the vicinity of X points, the dual-poloidal-region
$q$ coordinates   were developed in Sec. \ref{sec.3}. 
In extending the AEGIS to AEGIS-X from the surface-averaged $q$ to the dual-poloidal-region
 $q$ coordinates,  the following change is made 
\bean
\left. q(\chi) \right|_{AEGIS} \to \left. q_c(\chi)h(\chi,\theta) \right|_{AEGIS-X}
\eean
In AEGIS-X, the transformation of  $q_c(\chi)h(\chi,\theta)$ to a
matrix in the Fourier space is performed.  
 
In this formulation, the magnetic field is
expressed in \eq{bc0}, which is the complete representation. 
 Because of the dual-poloidal-region $q$ coordinates are used,
the difficulty of infinite resonance surfaces in the conventional surface-averaged $q$ description is avoided.

\begin{figure}[htp]\vspace*{-4mm}
\centering
\includegraphics[width=70mm]{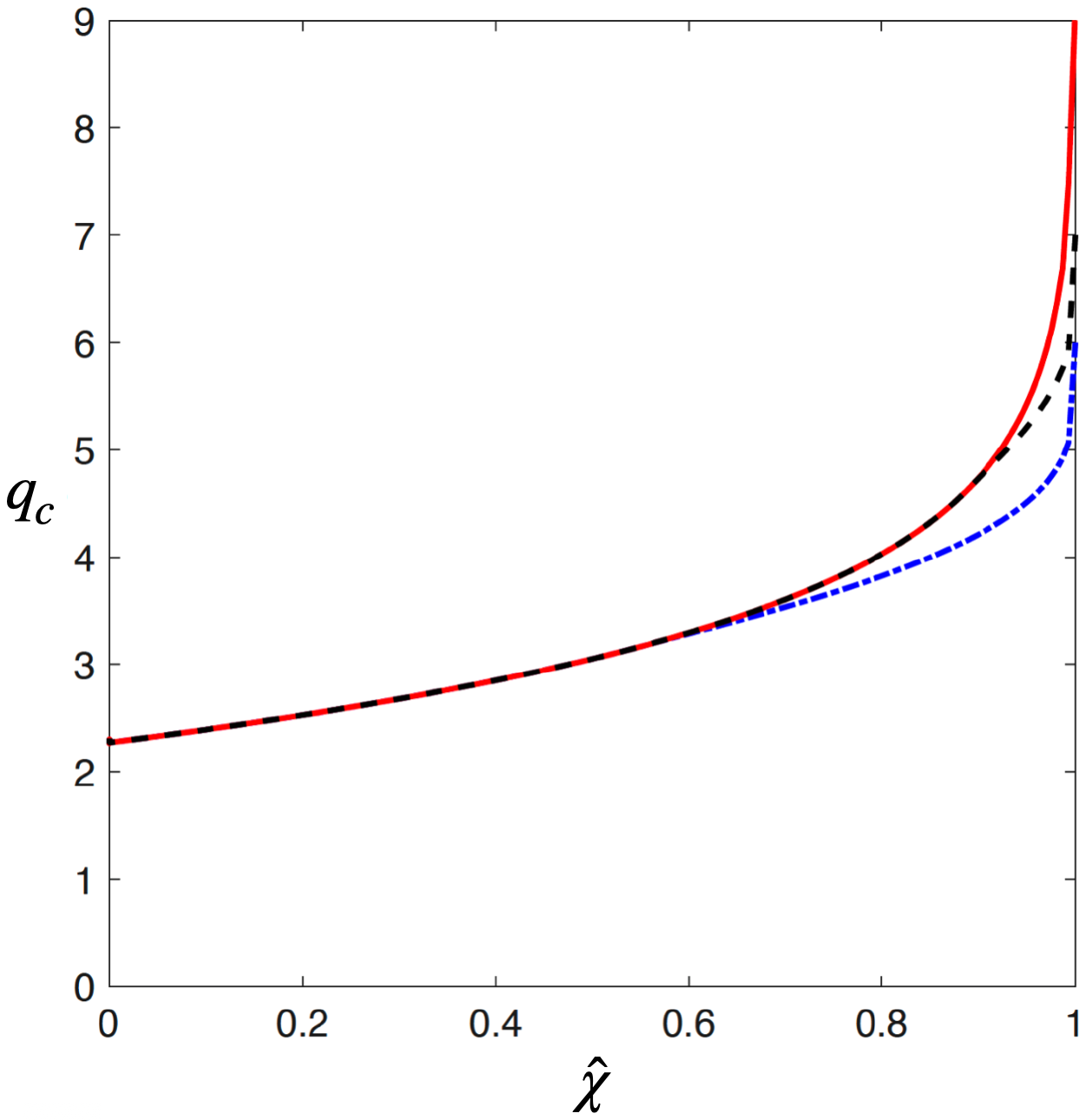}]\vspace*{-5mm}
\caption{The safety factor profiles  versus the normalized poloidal flux. 
The red curve corresponds to the surface-averaged $q$, which tends to infinity at $\hat \chi=1$,  the dashed and dot-dashed blue curves correspond
respectively to $q_c$ in the cases with $q_{c,a} = 7$ and $6$.}
\label{f8}
\end{figure}

\begin{figure}[htp]
\centering
\includegraphics[width=60mm,angle=-90]{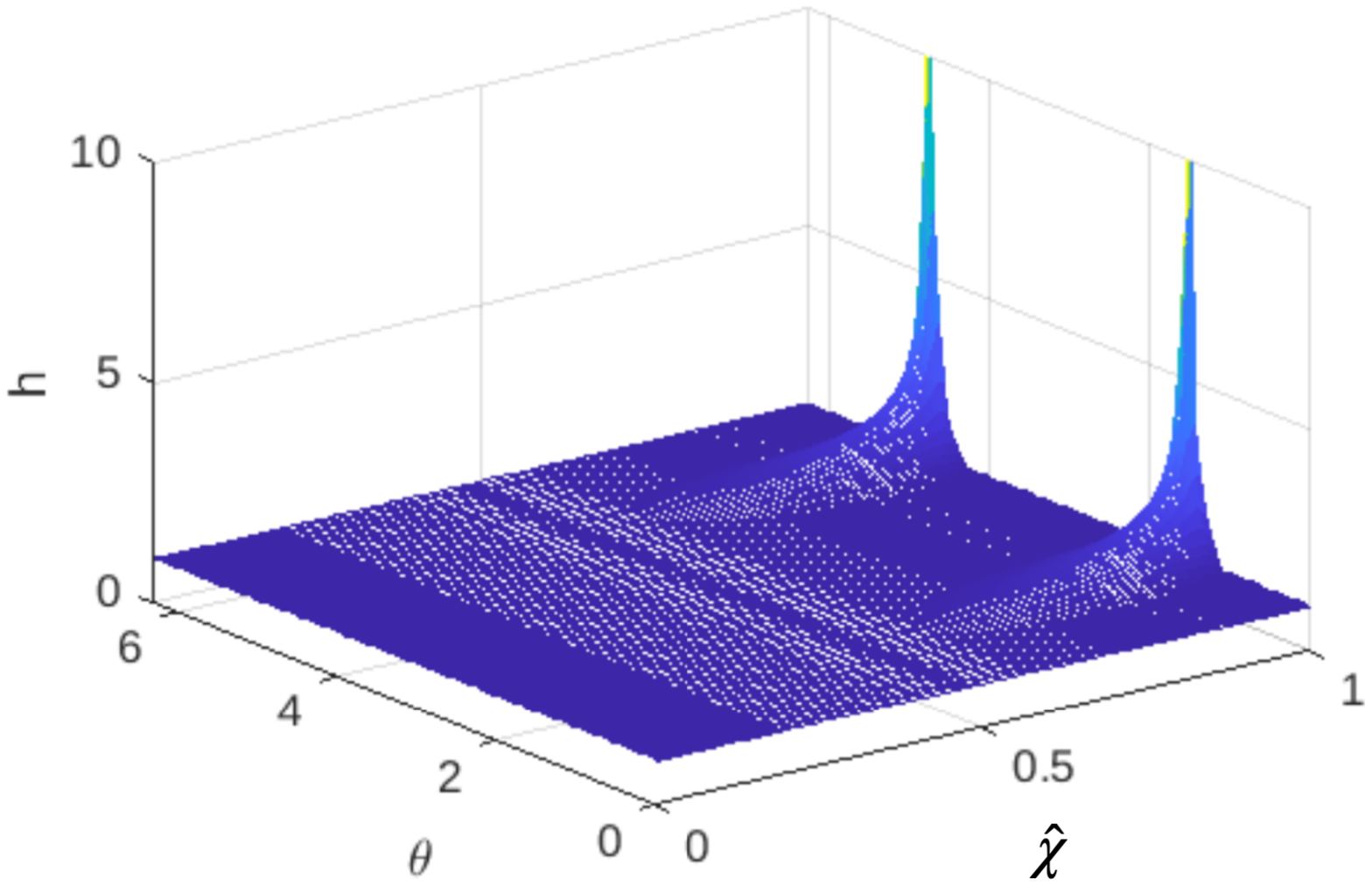}
\caption{The typical $h$ profile versus $\hat\chi$ and $\theta_c$. In this plot, $q_{c,a} = 6.8$ at $ \hat\chi=1$.}
\label{f9}
\end{figure}

We keep using the Solov\'ev equilibrium to demonstrate the X-point effects on the   external kink
modes. To be specific, we discuss the equilibrium with $\beta=0.03$, $X_0=3$, $X_s=2.33$, and $Z_s=1$
  as shown in Figs. \ref{f3}-\ref{f7}.
In the dual-poloidal-region $q$ coordinates, there is an assumption on where  the maximum Jacobian is imposed.
This is a parameter beyond what the MHD theory can fully determine. 
Note that the cutting-off of the maximum Jacobian corresponds to limiting the edge safety factor with the magnetic
field $\B_c$.  We, therefore, scan various cutting-off
positions, with $q_c$ at edge ($q_{c,a}$) ranging  from 5.8 to 7.2. The smaller $q_{c,a}$, the larger $\Theta_X$.
Figure \ref{f8} shows the surface-averaged $q$ profile 
together with the $q_c$ profiles for the cases with the edge safety factor 
$q_{c,a}=6.0$ and $7.0$ respectively. To see the safety factor feature,
the function   $h(\chi,\theta)$ for the case $q_{c,a} = 6.8$ is plotted in Fig. \ref{f9}.

\begin{figure}[htp]\vspace*{-22mm}
\centering
\includegraphics[width=85mm]{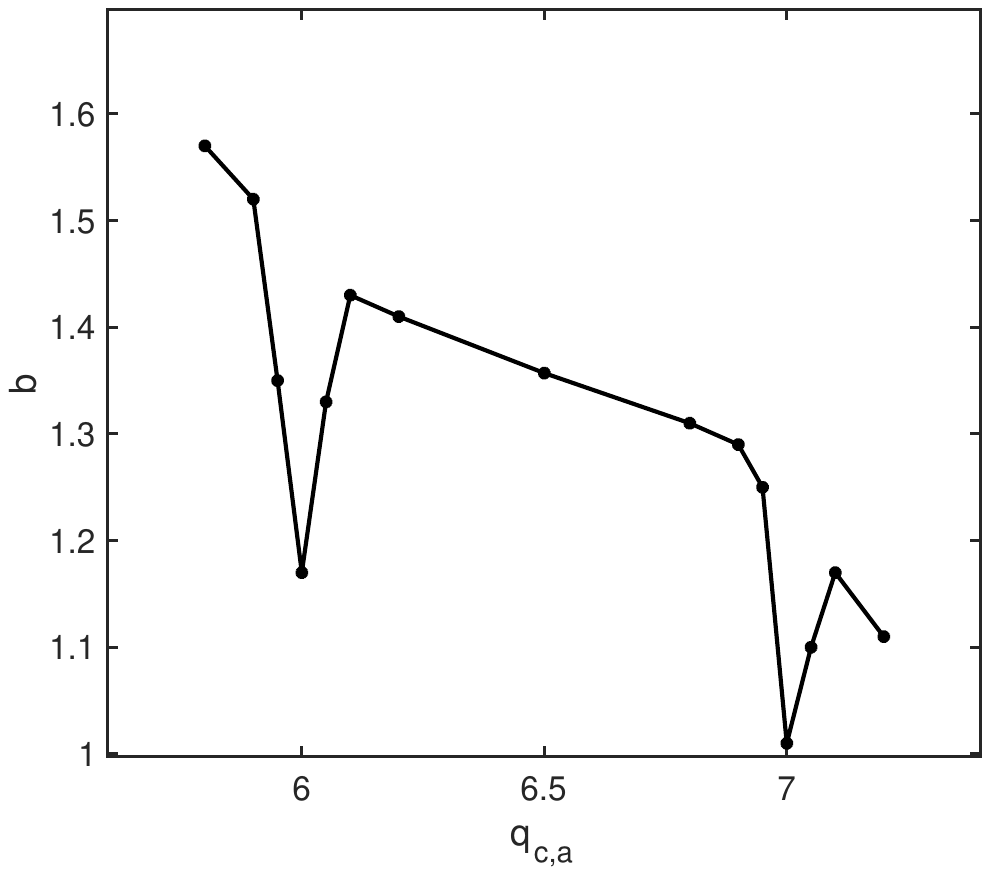}\vspace*{-21mm}
\caption{The critical wall positions versus the edge safety factor $q_{c,a}$   in the dual-poloidal-region coordinates.}
\label{f10}
\end{figure}

\begin{figure}[h]\vspace*{4mm}
\centering
\includegraphics[width=69mm,angle=-90]{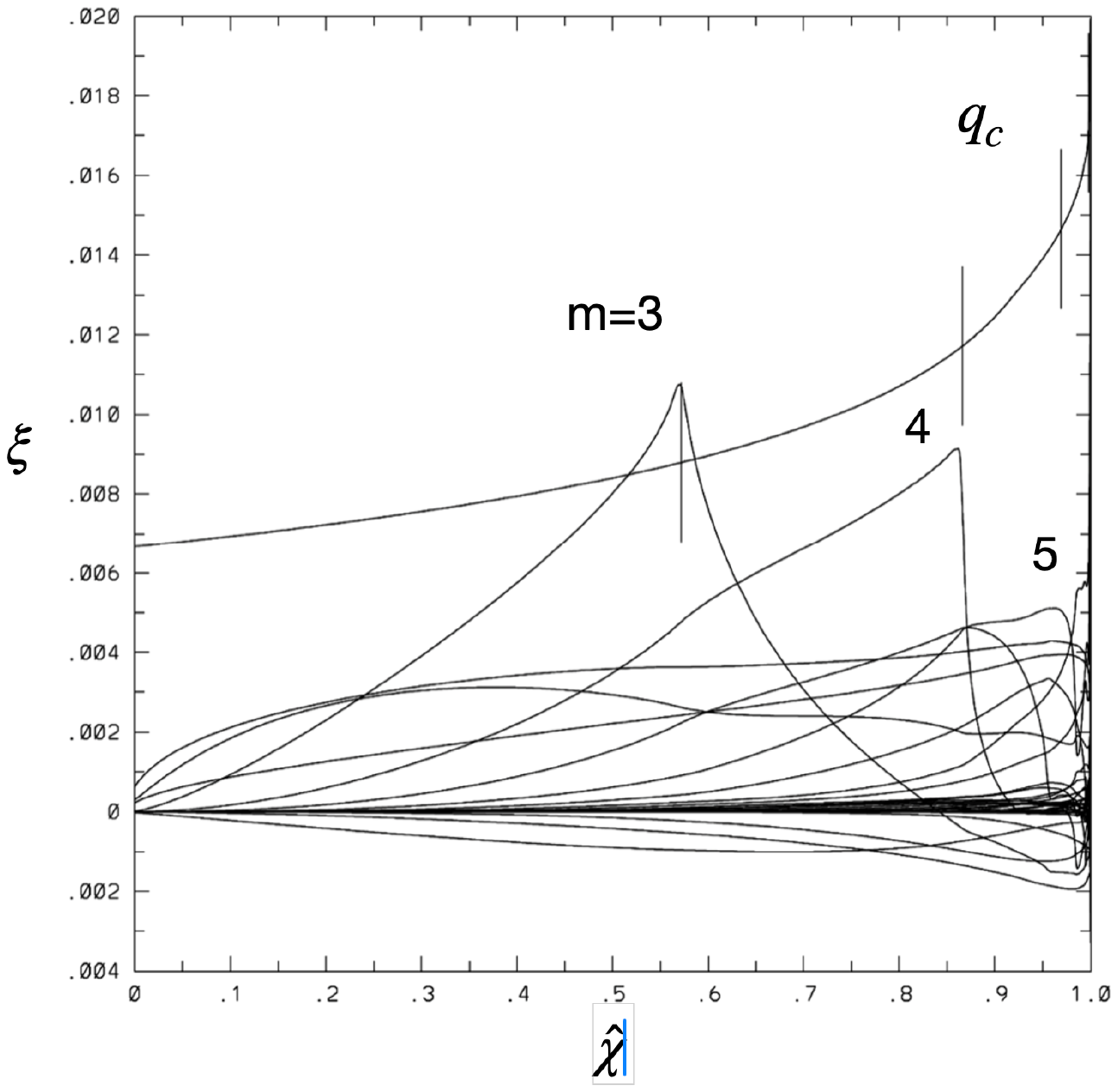}\vspace*{4mm}
\caption{The typical eigenfunction. The plot corresponds to the case with $q_{c,a}=6.8$.}
\label{f11}
\end{figure}

The $n=1$ modes are studied in the dual-poloidal-region $q$ coordinates. Fig. \ref{f10} shows the critical wall positions
for the cases with $q_{c,a} = 5.8, 6.0, 6.2, 6.8, 7.0$, and $7.2$ with the conformal wall.
The system is stable if the perfectly conducting wall is placed within the critical wall position,
otherwise unstable. Therefore, the system is more stable with a larger critical wall position.  
In the calculation, the coordinates   ($\chi,\theta_c,\phi$) with dual-poloidal-region $q$
are used. Therefore, we are able to study the modes aligned with the local magnetic field lines.
The existence of this type of modes is confirmed,  
The typical eigenmode in the dual-poloidal-region $q$ 
coordinates is given in Fig. \ref{f11} for the case with $q_{c,a}=6.8$.  

   From Fig. \ref{f10}  one can see that the larger $q_{c,a}$ in the overall trend 
the more unstable the system becomes. Note that the larger $q_{c,a}$ corresponds to
  the smaller $\Theta_X$. This indicates that
 the smaller the angle $\Theta_X$ in the overall trend, the more unstable the numerical results indicate.
This  is consistent with
the analytical theory for peeling modes in Sec. \ref{sec.4a}.

We have not reduced $\Theta_X$ further in the numerical calculation
 to determine the most stringent stability condition in the ideal MHD description.
As discussed earlier in Sec. III in discussing why we introduce
the dual $q$ coordinates, further reduction in $\Theta_X$ would place
the current treatment beyond the applicability limit of MHD description.   
In the poloidal direction, the distance between the
same magnetic field lines in the different toroidal loops,
$\delta L_X$ in Fig. \ref{f2} tends to be zero. In the radial direction, a big magnetic
shear appears. The kinetic effects become significant in the $\Theta_X$ region
for   external kink  modes. 
Furthermore, since the toroidal current density is constant
in the Solov\'ev equilibrium, which extends to the plasma edge, one needs 
infinitely fine grids or infinite number of poloidal
Fourier components  to minimize the field-line-bending energy,
which is beyond the numerical code capacity.  For these reasons,
we only display the tendency of stability criterion with respect to $\Theta_X$
at the current stage. We will consider this limit when full kinetic/resistive effects
are taken into account.

From Fig 9, one can also see  a typical feature of the peeling type of modes. The stability 
condition becomes more stringent whenever the resonance surface falls exactly on the plasma-vacuum interface. 
The critical wall position has a minimum when $q_{c,a}$ is an integer. Similar 
phenomena have been observed in the cases with the edge portion truncated.\cite{trun}

\section{Conclusions}

In this paper, the X-point effects on  the   ideal MHD modes in tokamaks 
are investigated using the dual-poloidal-region 
$q$ coordinates.     Since the X point effects mainly affects the edge region, 
the modes localized at tokamak edge are particularly examined. 
The procedure is an alternative to the finite element method used in this field.
 
Since the dual-poloidal-region safety factor is introduced, this makes
 our approach consistent with the  experimental observation that the filaments are aligned 
 with the local field lines in the  poloidally core region, $\Theta_{core}$. Indeed, from the analyses of the local safety factor
 using the Solov\'ev equilibrium,   it is confirmed  that the local safety factor actually tends to infinity
 only in the vicinity of X points, while it remains finite elsewhere. 
 Therefore, we conclude that the surface-averaged $q$  
 alone does not reflect the subtle feature of field line pitch.
 The coordinates based on the dual-poloidal-region $q$ can catch the   subtle  feature
 of the tokamak edge localized modes in the presence of X points. 

Two types of  magnetohydrodynamic  modes are studied using the dual-poloidal-region $q$ coordinates
both analytically and numerically.
Using the dual $q$ coordinates, we first prove that the X point effects can provide a stabilizing effect on
the    modes localized at tokamak edge by the singular mode theory,\cite{gre,glasser} 
which show that the smaller $\Theta_X$  the more unstable the system becomes. 
The results are confirmed in the AEGIS-X calculation.
We have not
considered the limit $\Theta_X \to 0$ in view of that the nonideal MHD effects can play a significant
role in the vicinity of X points.   
The existence of the modes aligned with the local magnetic field in the core region $\Theta_{core}$ is
confirmed. 
The analyses bear some kind of similarity with the conventional treatment with the surface-averaged
$q$ and with the edge portion truncated.

Here, we explain further that the X-point effects on the   external kink 
 modes cannot be fully resolved numerically in the ideal MHD description.
Neither our dual $q$ code  nor   existing ideal MHD  codes
can achieve the ``converged results" for X point effects in the ideal MHD framework
regardless of the numerical method used. 
The reasons are summarized here. 
  To minimize the stabilizing energy from the field line bending effects,  
the pressure gradient modes in the lowest order  tend to become flux-tube-like. 
The flux tube with  finite perpendicular size in the poloidally core region $\Theta_{core}$
 evolves into an infinitely thin sheet as it approaches the X points. 
Therefore, infinite fine mesh is required to resolve the flux tube modes
at the X points. This is not achievable numerically. Even if one could do it,
the perpendicular wavelength near the X points would exceed
 the ideal MHD applicable limit. The nonideal MHD effects, such as FLR
and resistive effects, can play significant roles. 
The dual $q$ description paves the way to include the nonideal MHD
effects by isolating the singularity to the vicinity of X points, which otherwise
spreads over the whole surface with infinite $q$ everywhere.

One can also see this from the derivation of Mercier's criterion. 
 Because the magnetic shear is infinite at the edge with X points, 
 the scale length of radial localized modes 
approaches  zero in the derivation of Mercier's criterion. 
This infinitely fine structure can be resolved analytically but not numerically.
This is why only the tendency of X point effects is shown in our  numerical work
in Fig. \ref{f10}. Instead, in the analytical
 peeling mode theory in  Sec. \ref{sec.4a}, one can take the limit
$\Theta_X=0$.

Using the dual-poloidal-region safety factor, we are able to determine   the condition of
the existence of the axisymmetric modes localized in the vicinity of X point.
The poloidal magnetic field vanishes in this region. The $n=0$ localized
modes can develop locally there   if the resistivity or external RMP drive 
is taken into account. This has important implications.
 More importantly, the existence of axisymmetric modes   (or the magnetic field pattern)
points to the possibility of applying a toroidally axisymmetric RMP in the X-point region for 
mitigating the edge localized modes. 
This   can be an alternative to  the current  RMP coil design in tokamaks. 
  Here, we would like to mention 
  the earlier research on the global vertical instabilities, for example,
in Refs.  \onlinecite{n01}-\onlinecite{n03}. The $n=0$ vertical modes are related 
to the low $m$ modes so that the plasma column moves vertically as a whole. 
The current $n=0$ axisymmetric  modes belong to the high $m$ case, which are localized
in the vicinity of X points. They are different. Nevertheless,
note that  in the investigation of X point effects on the global vertical instabilities, Refs.
 \onlinecite{n01} and \onlinecite{n02} pointed out that  the $n=0$ modes can have resonant 
effects at the X points. This can be a numerical support to the $n=0$ RMP concept pointed out in 
the current paper. 

The current dual $q$ description also has  important implications 
for the present concept of finite $m/n$ RMPs.
The  double-null (DN) configurations are thought to have better power handling 
and performance as compared to the single-null (SN) configurations,
However, the RMP ELM suppression  is usually only observed in the SN configurations
and there is no hint of RMP ELM suppression can be achieved in the DN configurations.\cite{rmpdn}
In the dual $q$ (or local $q$) picture, one can see the reason. if RMPs are configured according to
the averaged $q$, the deviation from the local $q$ in the DN configurations
is much larger than in the SN configurations. 
Taking into account the dual $q$ feature is also important for  
the existing concept of finite $m/n$ RMPs. It opens 
the possibility of further improving the current finite m/n RMP concept by considering
 the alignment to the local $q$.  Our dual $q$ description indicates
 that  both RMP concepts, finite $m/n$ and $n=0$, have their own
potential and deserve to be studied in parallel.

Nevertheless, this is the first effort to introduce the dual-poloidal-region $q$ description. 
Because of the variation of connection length and the dependence of finite Larmor radius effects
on the local safety factor, kinetic description\cite{tan,rev} may be needed to further clarify the edge
physics.   The resistivity effects can be also important in the vicinity of X points.  
We especially point out that in the ideal MHD description, $\Theta_X$ cannot be fully determined.
The applicability of ideal MHD theory shows that there is a lower limit for $\Theta_X$. For example,
$\delta L_x$ in Fig. \ref{f2} cannot be smaller than the ion Larmor radius  for MHD   to be physically relevant. 
  This is because for a flux-tube type of modes, for example the interchange or ballooning modes,
with a finite $\k_\bot$ in the region $\Theta_{core}$, their $\k_\bot$ in the region $\Theta_X$ would become
extremely large and therefore the FLR effects on them are important.
Since the complete determination of
$\Theta_X$ requires nonideal MHD theory,  in the current ideal MHD description,
we have,   nevertheless, performed a parameter scan of $\Theta_X$
and derived the  analytical  stability condition with $\Theta_X$ as a parameter
 to show the X-point stabilization. These are only for   external kink  modes.
The  nonideal MHD analyses of   external kink modes or peeling-ballooning type of modes are proposed for future studies.

Actually, this is not a special issue to the current dual-poloidal-region  $q$ description,
a similar or even more demanding requirement appears also in the single surface-averaged $q$ description.
The single surface-averaged $q$ description results in indefinitely dense singular surfaces at the edge.
Note that the distance of rational surfaces cannot be smaller than the Larmor radius for MHD   to be physically relevant. 
The single surface-averaged $q$ treatment meets the uncertainty for where to truncate the plasma edge portion
as pointed out in Ref. \onlinecite{trun}.
Both cases have a common origin from the X-point singularity. As a matter of fact, the required region for nonideal MHD
treatment is narrower in the dual-poloidal-region  $q$ description. 
In contrast,  the region is spread all over the magnetic surfaces near the edge in the single 
surface-averaged $q$ description.

  We also point out
the recent development of the X-point equilibrium research in Ref. \onlinecite{xeq}.
It is found that the X point actually does not exist on the plasma-vacuum interface,
but in the closely nearby vacuum region. The surface-average $q$ may not tend to be infinity
at the plasma edge. But, it is still large. To study the modes aligned with the local
magnetic field in the core region, the dual-poloidal-region $q$ description is still necessary in this case.

  The authors would like to acknowledge  Dr.
Richard Fitzpatrick for helpful discussion.
This research is supported by Department of Energy Grants DE-FG02-04ER54742.

\appendix

\section{Features of dual-poloidal-region $q$ coordinates}

\label{app.cood}

In this appendix, we discuss the features of dual-poloidal-region $q$ coordinates as described
in Sec. \ref{sec.3}.
\begin{figure}[h]\vspace*{-3mm}
\centering
\includegraphics[width=70mm,angle=-90]{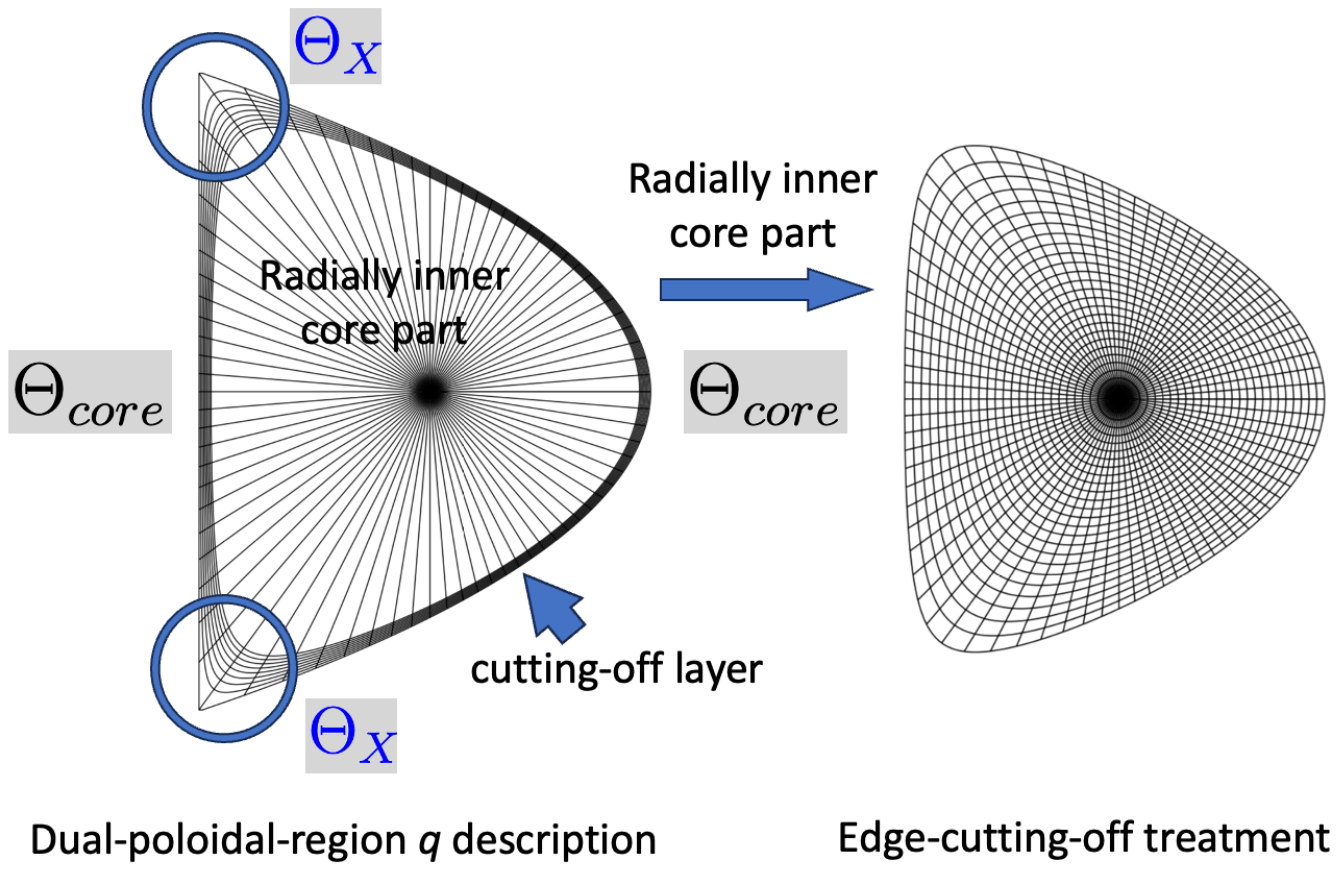} \vspace*{-10mm}
\caption{Comparison between the dual-poloidal-region $q$ description and the edge-cutting-off treatment.}
\label{fapp}
\end{figure}

In the single $q$ description with the conventional flux coordinates, the edge $q$ tends to be infinite. 
This leads considerable numerical codes based on this coordinate system to cut off some portion 
of edge region as shown in Fig. \ref{fapp}. Compared with this type of treatments, 
the dual-poloidal-region $q$ coordinate system reinstates the cutting-off edge region. 
In the dual-poloidal-region $q$ coordinate description,
the radially inner core part is described identically as the conventional treatment 
by the  flux coordinates with
single $q$, while the dual $q$ description is introduced for
the edge layer which is cut off in the conventional edge-cutting-off treatment. 

If one picked up the cutting-off layer in the single $q$ description, the
$q$ value in this layer would become infinite or rather larger. Instead, in the dual $q$ description,
the $q$ value remains finite in the poloidally core region, $\Theta_{core}$, and only becomes
infinite or rather large  in the vicinity of X points, $\Theta_X$.  

There are several advantages of dual-poloidal-region $q$ description. First, one can see
immediately that the localized axisymmetric mode may develop in  the vicinity of X points, $\Theta_X$.
Second,   the cutting-off treatment of external kink or peeling-ballooning modes tends to miss
the stabilizing effects from $\delta W_X$. Here, it has been noted that
   the cutting-off treatment of external kink or peeling-ballooning modes
 focuses on the modes with finite $m/n$
(i.e., around the edge safety value $q_a$),
which cannot align with the field lines in the vicinity of X points, $\Theta_X$.  
Furthermore, as pointed out in the main text, the X-point physics is intrinsically 
nonideal MHD. Isolating the singularity of $q_{local}$ in $\Theta_X$
is potentially helpful for nonideal MHD treatment of X-point physics. 
The ``singular point" theory may replace 
the conventional singular layer theory at the edge with X points.

\section{Ordering analyses of $n=0$ localized axisymmetric modes}

\label{sec.appendix}

In Sec. \ref{sec.4b}, we have derived the stability criterion of the localized
 axisymmetric modes, which in the ideal MHD is given by \eq{di4}. 
 The criterion shows that the balance between the pressure gradient drive
 and shear stabilization effects determines the stability. Note that in the resistive
 MHD, the $n=0$ localized axisymmetric modes are always unstable on the bad curvature
 side since the shear stabilization term disappears as proved in Ref. \onlinecite{glasser}.
 Here, we analyze the criterion in the ideal MHD case. 

Using the definitions in Sec. \ref{sec.3}, 
one can further reduce the first term on the right hand side of \eq{di4}.
Noting that
\bean
\B\bcdot \bnabla\theta_c &=& \bnabla\phi\btimes\bnabla\chi\bcdot\bnabla\theta_c
\nn
&=& \bnabla\phi\btimes\bnabla\chi\bcdot\bnabla\theta_{eq} \frac{\jac_{eq}^{\max} f}{X^2q_c}
\nn
&=&\jac_{eq}^{-1} \frac{\jac_{eq}^{\max} B}{Xq_c}
\eean
and
\bean
q_ch &=& q_c \frac{\jac_{eq} }{\jac_{eq} ^{\max}},
\eean
one has
\bea
\frac{g\mu_0P'\kappa_n}{(\B\cdot\nabla\Lambda_s)^2} &=&
\frac{ B^2 2 \mu_0 \Kb\bcdot\bnabla p}
{\lbs \B\bcdot\bnabla\theta   \bnabla\chi\bcdot \bnabla(q_ch)\rbs^2}
\nn
&=&  \frac{  (X_s^2 2 \mu_0/B^2) \Kb\bcdot\bnabla p }
{\lbm \lbs 1/B\rbs  \bnabla\chi\bcdot \bnabla \ln \jac_{eq} \rbm^2}.
\label{estim}
\eea
We now estimate the denominator of \eq{estim}.
Noting  that at the X point $\bnabla\chi= 0$,
one has   $\bnabla\chi= Ba\delta r/a$,
where $\delta r/a$ is the scaled distance from the X point.
Noting further  that $\jac_{eq}\sim Ra/|\bnabla\chi|\sim R/(B\delta r/a)$,
one obtains
\bea
\lbm \lbs 1/B\rbs  \bnabla\chi\bcdot \bnabla \ln \jac_{eq} \rbm^2
\sim (\delta r/\delta r)^2 \sim 1.
\label{bb2}
\eea
Therefore, by using Eqs. \eqn{estim} and \eqn{bb2},  the first term on the right-hand side of \eq{di4}  can be 
estimated as follows
\bea
\frac{g\mu_0P'\kappa_n}{(\B\cdot\nabla\Lambda_s)^2} \sim  (X_s^2 \mu_0/B^2) \Kb\bcdot\bnabla p.
\label{est1}
\eea
The stability condition in \eq{di4} indicates that the right-right side of \eq{est1} should be larger than
$1/4$ for the instability to occur  in the ideal MHD description.

\end{document}